\pdfoutput=1
\documentclass[twocolumn]{svjour3}          
\smartqed  
 
\usepackage{amsmath,amsfonts} 
\usepackage{algorithmic}
\usepackage{graphicx}
\usepackage{textcomp}
\usepackage[dvipsnames]{xcolor}
\usepackage{soul} 
\usepackage{amsfonts}
\usepackage{hyperref}
\hypersetup{colorlinks=false,%
            linkbordercolor=white,linkcolor=green,pdfborderstyle={/S/U/W 1}}

\usepackage[linesnumbered,ruled,vlined]{algorithm2e}
 
\usepackage{listings}

\usepackage{tikz}
\newcommand*\circled[1]{\tikz[baseline=(char.base)]{
            \node[shape=circle,fill,inner sep=1pt] (char) {\textcolor{white}{#1}};}}
\usepackage{makecell}

\newcommand{\tony}[1]{{\textcolor{red}{#1}}}
\newcommand{\myc}[1]{{\textcolor{black}{#1}}}

\newcommand{\change}[1]{{\textcolor{black}{#1}}}
\newcommand{\system}{\textsl{MorphStream}\xspace}
\newcommand{\stpg}{S\text{-}TPG\xspace}
\newcommand{\tpg}{TPG\xspace}

\newcommand{\blk}{$BLK$\xspace}
\newcommand{\rdy}{$RDY$\xspace}
\newcommand{\exe}{$EXE$\xspace}
\newcommand{\abt}{$ABT$\xspace}

\newcommand{\td}{\texttt{TD}\xspace}
\newcommand{\pd}{\texttt{PD}\xspace}
\newcommand{\ld}{\texttt{LD}\xspace}

\newcommand{\tds}{\texttt{TD}s\xspace}
\newcommand{\pds}{\texttt{PD}s\xspace}
\newcommand{\lds}{\texttt{LD}s\xspace}

\newcommand{\tstream}{TStream\xspace}
\newcommand{\sstore}{S-Store\xspace}

\newcommand{\se}{\hyperref[symbol:se]{\underline{\textsf{s-explore}}}\xspace}
\newcommand{\nse}{\hyperref[symbol:nse]{\underline{\textsf{ns-explore}}}\xspace}
\newcommand{\fsu}{\hyperref[symbol:fsu]{\underline{\textsf{f-schedule}}}\xspace}
\newcommand{\csu}{\hyperref[symbol:csu]{\underline{\textsf{c-schedule}}}\xspace}
\newcommand{\ea}{\hyperref[symbol:ea]{\underline{\textsf{e-abort}}}\xspace}
\newcommand{\la}{\hyperref[symbol:la]{\underline{\textsf{l-abort}}}\xspace}

\usepackage{enumitem}

\newenvironment{myenumerate}
{ \begin{enumerate}[leftmargin=0.2in]
		\vspace{-1ex}
		\setlength{\itemsep}{0pt}
		\setlength{\parskip}{0pt}
		\setlength{\parsep}{0pt}    }
	{ \end{enumerate}                  }
	
\usepackage{multirow}
\usepackage{subfig}

\usepackage{comment}

\usepackage{hyphenat}
\usepackage{url}

\newcommand{\compact}{\vspace{-5pt}}
\newcommand{\subcompact}{\vspace{-4pt}}

\usepackage[skip=3pt]{caption}
\usepackage{tcolorbox}
\usepackage{marginnote}
\newcommand{\margi}[1]{
\marginnote{
}
}

\newcommand{\margii}[2]{
\marginnote{
}
}
\usepackage[misc,geometry]{ifsym}

\begin{document}

\title{MorphStream: Scalable Processing of Transactions over Streams on Multicores}

\author{
Yancan Mao
\and Jianjun Zhao\thanks{Jianjun Zhao and Yancan Mao make equal contributions.}
\and Zhonghao Yang
\and Shuhao Zhang
\and Haikun Liu
\and Volker Markl
}

\institute{
              Yancan Mao \at
              National University Singapore, Singapore \\
              \email{maoyancan@u.nus.edu}             
              \and              
              Jianjun Zhao \at
              Huazhong University of Science and Technology, China \\
              \email{curry\_zhao@hust.edu.cn}             
              \and              
              Zhonghao Yang \at
              Singapore University of Technology and Design, Singapore \\
              \email{zhonghao\_yang@sutd.edu.sg}             
              \and              
              \Letter\  Shuhao Zhang \at
              Singapore University of Technology and Design, Singapore \\
              \email{shuhao\_zhang@sutd.edu.sg}             
              \and              
              Haikun Liu \at
              Huazhong University of Science and Technology, China \\
              \email{hkliu@hust.edu.cn}             \\      
              \and              
              Volker Markl \at
              Technische Universit{\"a}t Berlin, Germany \\
              \email{volker.markl@tu-berlin.de}           
}

\date{Received: date / Accepted: date}
\maketitle

\begin{abstract}
Transactional Stream Processing Engines (TSPEs) form the backbone of modern stream applications handling shared mutable states. 
Yet, the full potential of these systems, specifically in exploiting parallelism and implementing dynamic scheduling strategies, is largely unexplored. 
\myc{We present \system, a TSPE designed to optimize parallelism and performance for transactional stream processing on multicores. }
Through a unique three-stage execution paradigm (i.e., \textit{planning}, \textit{scheduling}, and \textit{execution}), \system enables dynamic scheduling and parallel processing in TSPEs. 
Our experiment showcased \system outperforms current TSPEs across various scenarios and offers support for windowed state transactions and non-deterministic state access, demonstrating its potential for broad applicability.
\end{abstract}

\section{Introduction}
\myc{The growing demand for real-time stream processing in data-intensive applications and the Internet of Things (IoT) has prompted the development of numerous Stream Processing Engines (SPEs)}, including Storm~\cite{storm}, Flink~\cite{flink}, and Spark-Streaming~\cite{spark}. 
\myc{
However, as stream applications grow and involve shared mutable states that are concurrently read and modified by multiple execution entities,
mainstream SPEs face significant challenges related to correctness~\cite{S-Store-demo} and efficiency~\cite{tstream}.
} 
To address these concerns, academia, and industry have turned their attention to TSPEs~\cite{tstream,Affetti:2017:FIS:3093742.3093929,S-Store,Transactions2018}, which offer built-in support for shared mutable states.

TSPEs employ transactional semantics to manage accesses to shared mutable states during continuous data stream processing. 
\myc{
Despite substantial efforts in the field, the current state-of-the-art TSPEs primarily rely on static task scheduling strategies, none of the existing TSPEs can maximize performance under different and dynamically changing workload characteristics, leaving a significant design space for scaling TSPEs on multicore processors largely unexplored. 
Additionally, due to complex control and data dependencies in workloads, existing TSPEs failed to fully leverage multicore parallelism.
}

\myc{
In this paper, we introduce \system, a novel TSPE designed to fill this gap. 
\system utilizes a Task Precedence Graph (\tpg)~\cite{hou1994genetic,669967,Robert2011,10.1145/167088.167254} that identifies the fine-grained dependencies among state access operations of a batch of state transactions. 
Based on \tpg, \system can schedule and execute concurrent state transactions dynamically.
To ensure both efficiency and correctness, \system further employs a three-stage execution paradigm, namely Planning, Scheduling, and Execution.
\begin{myenumerate}
\item \textit{Planning}: Based on a parallel two-phase \tpg construction process, \system efficiently tracks dependencies among transactions that may arrive out-of-order and involve special scenarios, i.e., access to non-deterministic states and windowed states.
\item \textit{Scheduling}: \system decomposes a scheduling strategy into three dimensions.
It dynamically adjusts scheduling decisions for \tpg on each dimension according to the input workloads and system state.
\item \textit{Execution}: \system correctly executes state transactions following scheduling decisions.
It is achieved by relying on the Stateful Task Precedence Graph (\stpg) to manage the lifecycles of transactions execution and the multi-versioning state table management to maintain the consistency and correctness of table entries.
\end{myenumerate}
}

\myc{
A preliminary version of this work was recently published in SIGMOD 2023~\cite{mao2023morphstream}, where we introduced the adaptive scheduling of concurrent state transactions in \system. 
}
Due to space constraints, many critical transaction execution details of \system were omitted. 
In this paper, we expand on those details and demonstrate the full breadth of \system's capabilities.
The advantages of \system extend beyond adaptive scheduling. 
\myc{
In this paper, we further discuss the scalability of \system on processing state transactions based on the three-stage execution paradigm. 
Additionally, we discuss how \system expertly integrates support for windowing operations and handles non-deterministic state access, amplifying its performance and adaptability even further. 
}

We experimentally demonstrate the capacity of \system to \myc{achieve} substantial improvements in throughput and latency for handling real-world use cases, compared to existing TSPEs. 
Furthermore, we show how to use \system to create two innovative applications, online social event detection~\cite{sahin2019streaming} and stock exchange analysis~\cite{sse}, which demonstrate \system's versatility and broad applicability. We open source the code, data, and scripts at \url{https://github.com/intellistream/MorphStream}.

\myc{
We have organized the rest of the paper as follows: 
Section~\ref{sec:background} provides a comprehensive background on transactional stream processing and delves into the design challenges inherent in optimizing a TSPE. 
Section~\ref{sec:Design_overview} presents a detailed overview of the design and execution workflow of \system, offering insights into its core functionalities.
Section~\ref{sec:planning} discusses how \system tracks dependencies in workloads and constructs \tpg. 
Section~\ref{sec:scheduling} discusses the adaptive scheduling strategies in \system.
Section~\ref{sec:execution} offers a detailed examination of transaction execution based on \stpg.
Section~\ref{sec:implement} showcases the programming model and APIs provided by \system, as well as the underlying system architecture that enables the three-stage execution paradigm.
Section~\ref{sec:exp} evaluates \system's performance using various benchmarks and real-world workloads. 
Section~\ref{sec:related} discusses related research to \system.
Finally, Section~\ref{sec:conclusion} concludes the paper and outlines future research directions.
}
\compact
\section{Preliminaries and Challenges}
\label{sec:background}
This section introduces the fundamental concepts of Transactional Stream Processing (TSP), elucidates the inherent limitations of existing TSPEs, and elaborates on the design challenges encountered in constructing an optimized TSPE. 

\subcompact
\subsection{Background}
We first provide the necessary background information on TSP, workload dependencies, and definitions related to state access operations, state transactions, and correct schedules.

\subsubsection{Transactional Stream Processing}
\label{subsubsec:tsp}
TSPEs differ from traditional SPEs, such as Storm~\cite{storm} and Flink~\cite{flink}, in their ability to maintain shared mutable states that can be referenced and updated by multiple execution entities (i.e., threads). These shared mutable states are preallocated in memory and expanded when needed before processing. We adopt the definitions from previous work~\cite{S-Store,tstream} for state access operations and state transactions, as well as the notion of a correct schedule for maintaining transactional semantics.

\paragraph{\textbf{State Access Operations:}}
A state access operation is a read or write operation to shared mutable states, denoted as \underline{$O_i$ = $Read_{t}(k)$ or $Write_{t}(k, v)$}. 
Timestamp ${t}$ is defined as the time of its triggering input event, while $k$ denotes the state to read or write that may be non-deterministic\footnote{See more discussions in Section~\ref{subsec:challenges}}, and $v$ denotes the value to write. 

\paragraph{\textbf{State Transaction:}}
The set of state access operations triggered by the processing of one input tuple is defined as one \emph{state transaction}, denoted as $txn_{t}$ = $< O_1, O_2, \ ... \ O_n >$. 
Operations of the same transaction have the same timestamp.
For brevity, we use the timestamp $t$ to differentiate different state transactions.

\paragraph{\textbf{Correct Schedule:}}
A schedule ($S$) of state transactions $txn_{t1}$, $txn_{t2}$, ..., $txn_{tn}$ is correct if it is \textbf{conflict equivalent} to $txn_{t1}$ $\prec$ $txn_{t2}$ $\prec$ $...$ $\prec$ $txn_{tn}$, where $\prec$ means that the left operand precedes the right operand.
\label{def:schedule}

\subsubsection{Workload Dependencies}
\label{subsec:definition}
\begin{figure}[t]
\centering
\includegraphics[width=0.48\textwidth]{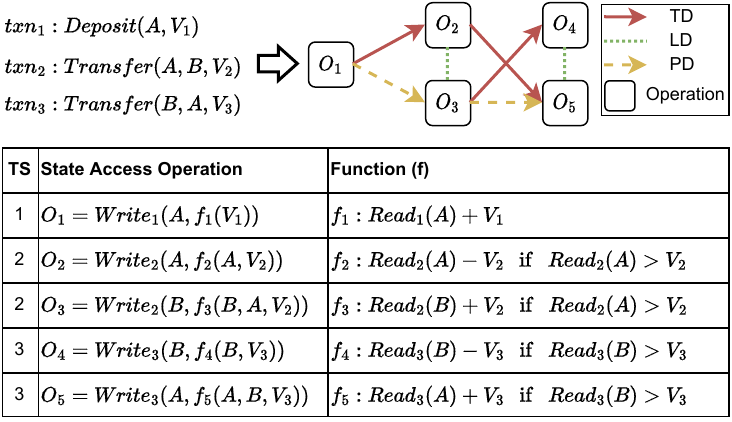}
\caption{Dependencies in Streaming Ledger (SL)}
\label{fig:dependencies_in_SL}
\end{figure}

To scale transactional stream processing, it is essential to maximize system concurrency while maintaining a correct schedule. However, this can be challenging due to the complex inter- and intra-dependencies among state transactions. By analyzing various TSPE applications~\cite{Arasu:2004:LRS:1316689.1316732,tstream,bidding,ACEP,Botan12,S-Store,Transactions2018}, we have identified three types of workload dependencies: logical, temporal, and parametric. These three types of dependencies play a crucial role in determining the scheduling of state transactions in TSPEs to maintain correctness and efficiency.
\myc{
We utilize the Streaming Ledger (SL) to demonstrate the dependencies depicted in Figure~\ref{fig:dependencies_in_SL}. 
SL represents a real-world stream application suggested by a recent commercial TSPE~\cite{Transactions2018} from data Artisans, and we examine three deposit/transfer transactions ($txn_1$, $txn_2$, and $txn_3$) within this context.
}

\paragraph{\textbf{Temporal Dependency (\td):}}
\label{def:TD}
\tds
are created when state access operations must comply with the event sequence. Let us denote two state access operations as $O_i$ and $O_j$. A temporal dependency exists between $O_i$ and $O_j$ if the following conditions are met: 
(1) Both $O_i$ and $O_j$ access the same state \emph{concurrently}; 
(2) The timestamp of $O_i$ is larger than the timestamp of $O_j$;
(3) $O_i$ and $O_j$ are not part of the same state transaction. 
In simpler terms, $O_i$ temporally depends on $O_j$ when they access the same state, but $O_i$ has a later timestamp, and they belong to separate state transactions.

\paragraph{\textbf{Parametric Dependency (\pd):}}
\label{def:PD}
\pds
arise between two write operations if the value to be written in one operation is contingent upon the execution of another operation. Let us denote two write operations as $O_i=Write(k_i,v)$ and $O_j=Write(k_j, v')$ where $v = f(k_1, k_2, ..., k_m)$. A parametric dependency exists between $O_i$ and $O_j$ if the following conditions are met: 
(1) $k_j \neq k_i$;
(2) $k_j$ is included in the set ${k_1, k_2, ..., k_m}$;
(3) The timestamp of $O_i$ is larger than the timestamp of $O_j$. 
In other words, $O_i$ parametrically depends on $O_j$ when $O_i$'s write value is determined by the result of $O_j$, and $O_i$ has a later timestamp.

\paragraph{\textbf{Logical Dependency (\ld):}}
\label{def:LD}
\lds
are essential in maintaining ACID properties. They stipulate that the aborting of a single operation necessitates the aborting of all operations within the same state transaction. In other words, operations $O_i$ and $O_j$ are considered to exhibit a logical dependency if they belong to the same state transaction. During the execution phase, operations can be scheduled without considering logical dependencies. However, if an operation $O_i$ is to be aborted, all other operations within the same transaction will also be aborted. This is in order to uphold the ACID properties by ensuring that aborting one operation results in aborting all operations of the same state transaction.

\subcompact
\subsection{Limitations of Existing TSPEs}
\label{subsec:related_work}

In this subsection, we review the related work on TSPEs, focusing on two state-of-the-art TSPEs: S-Store and TStream.

\textbf{S-Store}~\cite{S-Store} splits shared mutable states into disjoint subsets called partitions. It schedules each state transaction as the unit of scheduling. S-Store enforces all three types of workload dependencies: temporal, parametric, and logical. State transactions with contended state accesses are serially executed in timestamp order, preserving temporal and parametric dependencies. Operations within the same transaction are also executed serially, preserving logical dependencies. This scheduling scheme has several advantages. It is easy to implement, minimizes context-switching overhead, and performs well when state transactions do not overlap. However, it offers limited parallelism when state transactions overlap, leading to reduced system performance.

\textbf{TStream}~\cite{tstream} partitions state transactions into atomic state access operations and assembles operations targeting the same state into timestamp-sorted groups called operation chains. These operation chains are executed in parallel as long as there are no \myc{parametric} dependencies among them. 
The remaining operation chains to process are left to the operating system.
TStream's scheduling strategy is driven by the need to enforce temporal dependencies. However, it may still suffer from random blocking (i.e., busy waiting) due to potentially unsolved parametric dependencies among operation chains. 
Furthermore, TStream overlooks logical dependencies, and aborts can only be handled when the current batch of state transactions is fully processed. 
This leads to significant wasted computation and costly rollback operations.

\emph{In summary}, existing approaches to TSPEs either actively track all types of dependencies with a serial schedule or implicitly resolve certain dependencies with locks. Both S-Store and TStream have their limitations in terms of parallelism and abort handling. There is a need for \myc{an optimized TSPE with improved scheduling strategy and execution paradigm} that can maximize system concurrency while maintaining correct schedules and efficiently handling aborts.

\subcompact
\subsection{Design Challenges}
\label{subsec:challenges}
Maintaining correctness and efficiency in TSPEs hinges on effectively addressing a host of challenges. These challenges span across dependency identification, state transaction scheduling, and operation execution. We identify three significant challenges:

\paragraph{C1: Out-of-order Data Arrival:}
Stream data processing is complicated by the unpredictable arrival order of data, which poses challenges in accurately identifying dependencies. Specifically, when the order of state transactions fails to align with their timestamps, potential discrepancies can emerge. For example, consider two state transactions, $txn_{1}$ and $txn_{2}$, as illustrated in Figure~\ref{fig:dependencies_in_SL}. While $txn_{2}$'s higher timestamp suggests that $O_{2}$ should depend temporally on $O_{1}$, the out-of-order arrival (i.e., $txn_{1}$ arriving after $txn_{2}$) can create a paradox. Mitigating this challenge necessitates TSPEs to develop mechanisms that accurately capture dependencies, regardless of the state transaction's arrival sequence.

\paragraph{C2: Non-deterministic State Access:}
For the correct execution of state transactions, it is essential to identify all dependencies before scheduling state access operations, implying that transactions should be aware in advance of all states they will interact with. 
\myc{However, when state access is non-deterministic, meaning the accessed state is influenced by factors beyond the input event~\cite{Clonos}, immediate dependency determination becomes challenging.}
This is further complicated in the scenario of range access, where an event triggers interaction with an arbitrary number of states, requiring precise tracking, ordering, and execution of the access operations for each state. Additionally, in the event of an abort, comprehensive rollback of all changes associated with the range access is critical. To address these challenges, TSPEs need to judiciously manage state access and ensure deterministic execution (Section~\ref{def:schedule}).

\paragraph{C3: Window Operations:}
Window operations~\cite{golab2}, which create temporal boundaries for data aggregation and processing, add further complexity to dependency identification and scheduling. 
Dependencies may span across multiple windows, each of which can vary in size and can overlap with others, further complicating dependency tracking. 
Sliding windows, which shift continuously over time, introduce an additional layer of complexity. 
\myc{
Concurrent execution of multiple window operations introduces complexities, as each operation requires access to specific states and collects input data within a designated window range. 
This necessitates the implementation of sophisticated state management techniques to ensure accurate processing of window operations.
}
\compact
\section{Design of MorphStream}
\label{sec:Design_overview}
In this section, we delineate the design principles underpinning \system and its execution workflow.

\subsection{Design Principles}
The central principle of \system, designed to address the challenges outlined earlier, involves mapping the state transaction scheduling problem onto a \tpg scheduling problem. \system constructs a \tpg such that each vertex corresponds one-to-one with an operation. The dependencies among operations, as defined in Section~\ref{def:TD}, are then directly mapped onto edges connecting these vertices. Thus, processing all operations of a list of state transactions $L$, while respecting these dependencies, ensures a correct schedule. This concept is embodied in a \tpg, where, for example, operations $O_1$ $\sim$ $O_5$ in Figure~\ref{fig:dependencies_in_SL} and their dependencies naturally form a \tpg. Based on this principle, \system handles dependency identification and state transaction scheduling challenges via the following designs:

\paragraph{D1: Batched and Sorted Transaction Processing:} \system employs a two-phase \tpg construction process (Section~\ref{sec:planning}). The initial phase sorts the batched, out-of-order state transactions, while the second phase identifies dependencies based on the sorted transactions.

\paragraph{D2: Virtual Operation Implementation:} \system introduces \emph{virtual operation} (Section~\ref{sec:planning}) in the \tpg to handle non-deterministic state access. It tracks potential dependencies and helps anticipate state access needs. Furthermore, \system records states accessed by non-deterministic transactions after execution (Section~\ref{sec:execution}) to ensure accurate rollback when necessary.

\paragraph{D3: Generalized Structure for Window Operations:} \system addresses complexities introduced by window operations by generalizing their structure. By comparing overlapping windows (Section~\ref{sec:planning}), \system identifies dependencies among these operations. Furthermore, it retrieves records from the multi-versioning table within the window boundaries (Section~\ref{sec:execution}), ensuring accurate data processing within each window.

\subsection{Three-stage Execution Paradigm}
\label{subsec:overview}

\begin{figure}[t]
\centering
\includegraphics[width=0.48\textwidth]{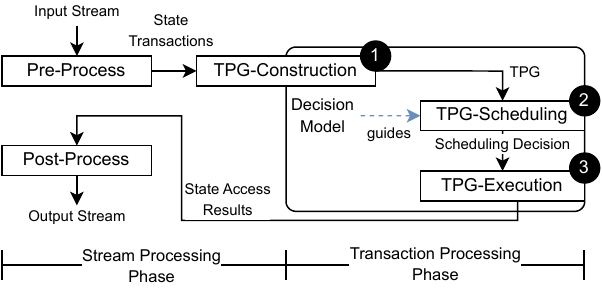}
\caption{The execution workflow of \system.}
\label{fig:workflow}
\end{figure}

As depicted in Figure~\ref{fig:workflow}, \myc{\system executes concurrent state transactions} in three stages, each contributing to its efficient and accurate operation:


\textbf{\circled{1}} \textit{Planning:} \system constructs the \tpg by identifying fine-grained temporal, logical, and parametric dependencies within and among a batch of state transactions. This involves implementing the design strategies outlined above, including the two-phase \tpg construction process, the use of virtual operations, and the application of specific dependency tracking rules for window operations.

\textbf{\circled{2}} \textit{Scheduling:} operations are scheduled for execution based on the \tpg (Section~\ref{sec:scheduling}). A decision model (Section~\ref{subsec:model}) guides \system in making optimal scheduling decisions, considering varying workload characteristics.

\textbf{\circled{3}} \textit{Execution:} Threads execute operations concurrently based on the scheduling decisions, while ensuring state access correctness. \system employs a finite state machine for each operation to accurately capture its state access behavior during execution and aborting. It relies on the multi-versioning state table management to \myc{maintain the correctness and consistency of table entries}. \system supports window-based state access by querying a range of targeting state copies from the multi-versioning state table. Furthermore, it records state accesses for non-deterministic state transactions after execution to ensure accurate rollback if necessary.



\compact
\section{Planning: \tpg Construction and Dependency Tracking}
\label{sec:planning}
In this section, we detail the process of \tpg construction along with how \system tracks dependencies in workloads.

\subcompact
\subsection{Planning Overview}
\label{subsec:construction}

\begin{figure*}[t]
\centering
\includegraphics[width=\textwidth]{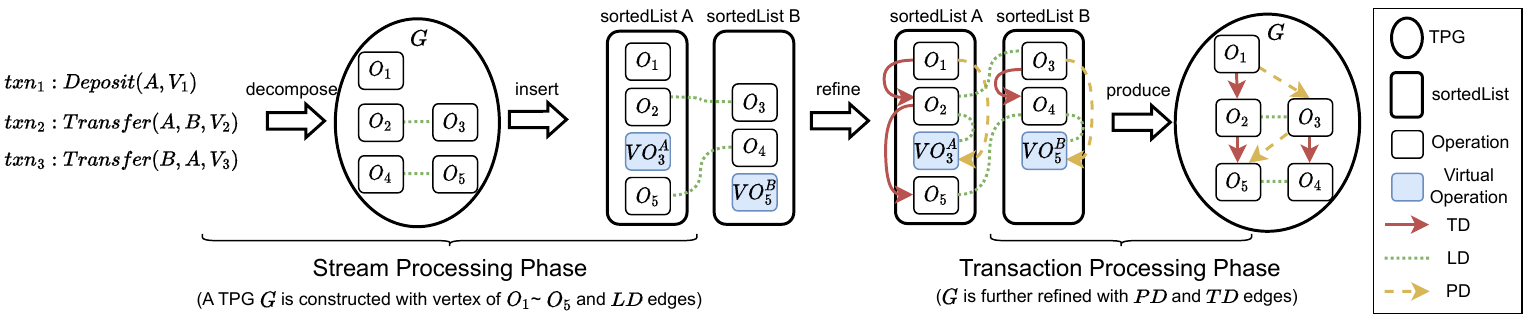}
\caption{A running example of a \tpg construction process involving three state transactions, which may arrive out-of-order.}
\label{fig:TPG}
\end{figure*}

\system constructs a \tpg for every batch of state transactions with minimal overhead. Upon the arrival of transactions, they are decomposed into atomic state access operations, which then serve as the vertices of the \tpg. Two main techniques are employed by \system to aid in the construction of the \tpg: \emph{Two-phase \tpg construction} and \emph{Virtual Operations}.

\textbf{Two-phase \tpg construction:} 
To counter the issue of out-of-order state transactions, \system partitions the \tpg construction process into two phases: stream processing phase and transaction processing phase. During the stream processing phase, transactions are decomposed into atomic state operations, which are arranged as per the temporal order of the transactions and serve as vertices of the \tpg. During the transaction processing phase, three types of dependencies - 
\myc{\ld, \td and \pd}
- are identified among these state operations to construct the \tpg.

\textbf{Virtual Operations:} To identify dependencies for an operation that accesses multiple \myc{(e.g., $n$)} states, $n$ virtual operations of the operation are maintained. Each virtual operation represents one state access on a specific state and is inserted into sorted lists of corresponding states. This method allows \system to efficiently identify dependencies during the \tpg construction process.

\subcompact
\subsection{Dependency Tracking for Out-of-Order State Transactions}
\label{subsec:out-of-order}
The construction of the \tpg includes the identification of \lds among operations from the same transaction. However, due to the potential out-of-order arrival of transactions, \tds and \pds cannot be immediately identified. To solve this issue, \system utilizes two-phase \tpg construction and virtual operations to identify all three types of dependencies:

\textbf{Stream processing phase:} Transactions are decomposed into atomic state access operations and \lds are identified. A preliminary \tpg is constructed by inserting operations as vertices and \lds as edges. To identify \tds, operations are inserted into key-partitioned sorted lists (based on the targeted state of each operation) sorted by timestamp. For each write operation with multiple states, virtual operations are maintained and inserted into the sorted lists to identify \pds during the next phase.

\textbf{Transaction processing phase:} At this phase, further state transactions are blocked until the stream processing phase resumes. \tds and \pds can be identified efficiently using the constructed sorted list and virtual operations from the stream processing phase. \tds are identified by iterating through operations in each sorted list. \pds are identified based on the precedence of the virtual operations in the sorted list. After identifying all \tds and \pds, these are inserted as edges into the \tpg.

\textbf{Running Example.}
\myc{A practical application of this process is in SL, which involves two types of transactions: deposit transactions and transfer transactions.}
Figure~\ref{fig:TPG} shows an example involving a deposit transaction and two transfer transactions that arrive consecutively. Through the two-phase \tpg construction process and the use of virtual operations, the dependencies are tracked efficiently, and a comprehensive \tpg is constructed.

\myc{
In the stream processing phase, upon arrival, transactions $txn_1$, $txn_2$, and $txn_3$ are decomposed into atomic state access operations $O_1$, $O_2$, $O_3$, $O_4$, and $O_5$. 
$txn_1$ is decomposed into $O_1$, $txn_2$ into $O_2$ and $O_3$, and $txn_3$ into $O_4$ and $O_5$. 
}
\lds among $O_2, O_3$ and $O_4, O_5$ are identified as they are from the same transaction and are ordered by their statement orders. A preliminary \tpg is constructed by inserting these operations as vertices and the \lds as edges. 
\myc{
The operations are then inserted into two sorted lists (one for each state, A and B). For operations $O_3$ and $O_5$, which have write functions dependent on states $A$ or $B$, virtual operations ($VO_3^{A}$ and $VO_5^{B}$) are inserted into the sorted lists of states $A$ and $B$, respectively.
}

During the transaction processing phase, \tds can be identified among $O_1, O_2, O_5$ and $O_3, O_4$ by iterating through the operations in each sorted list. \pds are identified between $O_1$ and $VO_3^{A}$, and between $O_3$ and $VO_5^{B}$ according to the previously inserted virtual operations. After the identification of \tds and \pds, they are inserted as edges to refine the preliminary \tpg to become the final \tpg of the current batch of state transactions.

\subcompact
\subsection{Dependency Tracking for Window Operations}

\begin{figure}
\begin{center}
	\begin{minipage}{0.49\textwidth}
        \subfloat[Window.]{   
		\includegraphics[width=0.48\textwidth]
		{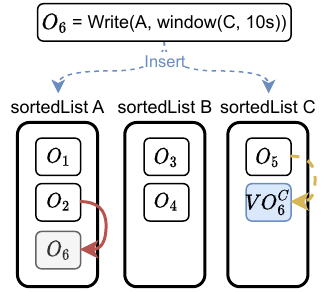}   
		\label{fig:windowoperation}
	}
        \subfloat[Non-deterministic.]{
		\includegraphics[width=0.48\textwidth]{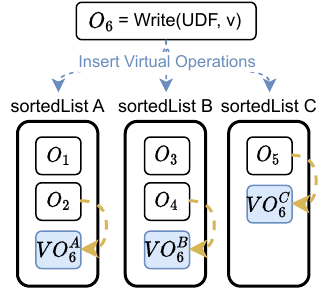}
        \label{fig:nondeterminisitc}
	}
	\end{minipage}
	\caption{Dependencies tracking for two special cases.}
    \label{figures:special_cases}
\end{center}    
\end{figure}

Window operations add another layer of complexity to dependency tracking as these operations may operate on a range of states defined by the window size. To manage this complexity, we present a generalized structure of window operations and identify dependencies among these operations by comparing their overlapping windows. In alignment with previous work~\cite{golab2}, a window operation in \system is modeled as a read or write operation, further associated with a time range and a trigger time. 

Dependencies among window operations are determined based on the trigger time and time range, enabling the identification of conflicting read or write operations. This facilitates the construction of \tpg and the scheduling of window operations. Specifically, a window operation is temporally or parametrically dependent on another operation if: 1) it overlaps with another operation in terms of state access, and 2) its window trigger happens after the trigger of another operation, and their window time ranges overlap. The dependency tracking for window operations follows a similar process as for normal operations, identifying dependencies based on the keys accessed and window trigger time. This approach works because two successive operations always share an overlapping window range.

A running example of dependency tracking among window operations is provided in Figure~\ref{fig:windowoperation}. The objective, similar to that in Section~\ref{subsec:nondeterminisitc_dependency}, is to identify dependencies for a new non-deterministic state access operation $O_6$ among five operations in three $sortedLists$ $A, B, C$. Specifically, the operation is a window write operation, tasked to aggregate state access of $C$ within the past 10 seconds and write the results to the target state $A$. Following the dependency tracking mechanism, we insert operation $O_6$ into the $sortedList$ $A$ and a virtual operation $VO_6^{C}$ into the $sortedList$ $C$. Subsequently, we establish a \pd between $O_5$ and $O_6$, and a \td between $O_2$ and $O_6$.

\subcompact
\subsection{Dependency Tracking for Non-deterministic State Operations}
\label{subsec:nondeterminisitc_dependency}

Non-deterministic state operations pose a challenge for dependency tracking as the states these operations access remain undetermined initially. In other words, the key $k$, written by an operation as discussed in Section~\ref{subsubsec:tsp}, can itself be a function. Therefore, the state to read or write depends on the result of the function evaluation. However, a pessimistic assumption can be made that in the worst-case scenario, a non-deterministic state operation could potentially access all states. Consequently, this operation is considered dependent on preceding operations of all states. 

To facilitate this, we introduce virtual operations of a non-deterministic state operation into all possible states during \tpg construction, establishing potential parametric dependencies with other operations. Particularly, \system tracks \pds for non-deterministic state transactions by leveraging virtual operations, following an approach akin to the one described in Section~\ref{subsec:out-of-order}. This approach permits \system to accurately schedule non-deterministic state transactions, albeit at the cost of potential parallel execution opportunities.

An illustrative example showcased in Figure~\ref{fig:nondeterminisitc}, elucidates dependency tracking for non-deterministic state operations. Initially, five atomic state access operations $O_1 \sim O_5$ are placed in three $sortedLists$ $A, B, C$. The objective is to ascertain dependencies for a newly introduced non-deterministic state access operation $O_6$. Specifically, $O_6$ writes a value $v$ to a state determined by a user-defined function $UDF$. By applying the dependency tracking mechanism, we incorporate three \emph{virtual operations} $VO_6^{A}, VO_6^{B}, VO_6^{C}$ into the three $sortedLists$ $A, B, C$. Subsequently, \pds for $O_6$ are identified in each $sortedList$ accordingly.

\compact
\section{Scheduling: \tpg-Based Decision-Making}
\label{sec:scheduling}
This section details the scheduling strategy of \system, a dynamic, multi-dimensional approach that capitalizes on the \tpg for efficient and correct stream processing task execution. 

The \system scheduling strategy operates on three dimensions.
\textit{Exploration Strategies:} These define how the \tpg is traversed for scheduling operations. The goal is to balance the depth and breadth of exploration for optimizing the concurrency-overhead trade-off (Section~\ref{subsec:explore}).
\textit{Scheduling Unit Granularities:} These pertain to the task size that is scheduled, which can range from individual state access operations to entire state transactions. Finer granular units facilitate more parallelism, potentially at the expense of increased overhead (Section~\ref{subsec:granularity}).
\textit{Abort Handling Mechanisms:} These strategies manage transaction aborts, affecting overall system performance and consistency (Section~\ref{subsec:abort_handling}).

\begin{table*}[t]
\centering
\caption{Scheduling decisions at three dimensions}
\label{tab:decisions}
\resizebox{0.99\textwidth}{!}{%
\begin{tabular}{|p{3.5cm}|p{1.5cm}|p{5.5cm}|p{5.5cm}|}
\hline
\textbf{Dimension}                                      & \textbf{Decision} & \textbf{Pros} & \textbf{Cons} \\ \hline
Exploration Strategy        & \se      &    \change{Threads can run in parallel with minimum coordination}  &  \change{BFS: Sensitive to workload imbalance / DFS: High memory access overhead}   \\ \cline{2-4} 
                                               & \nse     &   \change{More parallelism opportunities}   &   \change{Higher message-passing overhead}   \\ \hline
Scheduling Granularity & \fsu     &    \change{Better system scalability}  &   \change{High context switching overhead}  \\ \cline{2-4} 
                                               & \csu     &   \change{Lower context switching overhead}   &    \change{Less scalable and more sensitive to load imbalance}  \\ \hline
Abort Handling     & \ea      &    \change{Less wasted computing efforts}   &  \change{High context switching overhead}   \\ \cline{2-4} 
                                               & \la      &  \change{Less context switching overhead}    &   \change{More wasted computing efforts}   \\ \hline
\end{tabular}%
}
\end{table*}

Each of these dimensions assists \system in adapting to different workload patterns and system states. Table~\ref{tab:decisions} summarizes potential decisions within each dimension.
To optimize these dynamic scheduling decisions, we further introduce a heuristic decision model. This model evaluates the current workload characteristics at runtime, enabling appropriate scheduling decisions (Section~\ref{subsec:model}).

\begin{figure}[t]
\centering
\includegraphics[width=0.48\textwidth]{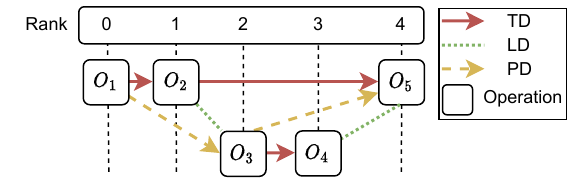}
\caption{A stratified auxiliary structure of \tpg.}
\label{fig:structured}
\end{figure}

\subcompact
\subsection{Exploration Strategies}
\label{subsec:explore}
Threads traverse the \tpg to select operations for processing, employing either structured or unstructured exploration strategies.

\label{symbol:se}
In the structured approach (\se), threads adopt depth-first or breadth-first traversal patterns. As per Nikolov et al.'s work~\cite{nikolov2006graph}, \se involves partitioning vertices into subsets based on connected directed paths, with subsets assigned ranks that inform stratum placement. Two methods stem from this approach.

\emph{A) BFS-like Exploration:} Threads concurrently explore operations of the same stratum, advancing to the next stratum only once all operations within the current one have been processed. This method relies heavily on barrier-based synchronization and its effectiveness can be compromised by uneven workload distribution among threads due to unpredictable operation completion times and potential aborts.
\emph{B) DFS-like Exploration:} Here, threads are initially assigned an equal number of operations across strata. A thread can advance to the next stratum once the dependencies of its assigned operations in the current stratum are resolved. This method necessitates less synchronization overhead, as threads progress based on their assigned operations rather than waiting for all threads to complete a stratum. This, however, may increase memory access overhead due to repeated dependency resolution checks.

\label{symbol:nse}
Alternatively, in the non-structured approach (\nse), threads randomly select operations whose dependencies are resolved. Each thread maintains a signal holder that asynchronously manages dependency resolutions for remaining operations. Specifically, when an operation $O_i$ is successfully processed, the thread signals all other threads, which can then process operations dependent on $O_i$. Despite ensuring immediate resolution of operation dependencies and making more operations available for scheduling, \nse can lead to higher message-passing overhead due to the need for countdown latch signal transmission along all directed edges.

\begin{figure}[t]
\centering
\includegraphics[width=0.48\textwidth]{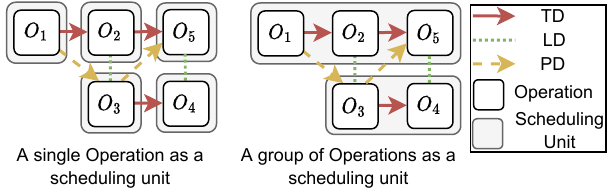}
\caption{Different scheduling unit granularities.}
\label{fig:granularity}
\end{figure}

\subcompact
\subsection{Scheduling Unit Granularities}
\label{subsec:granularity}
\system offers flexibility in scheduling granularity, allowing either single-operation (Fine-grained Scheduling Unit or \fsu) or multi-operation (Coarse-grained Scheduling Unit or \csu) scheduling. 
\label{symbol:fsu} In \fsu mode, a thread schedules single operations, as depicted in Figure~\ref{fig:granularity}(a). This approach optimizes system scalability by maximizing opportunities for parallelism and enabling immediate dependency resolution. 

\label{symbol:csu}
The drawback of \fsu
is that it incurs significant context-switching overhead. Conversely, \csu mode schedules groups of operations, which helps reduce context-switching overhead at the potential expense of slowing dependency resolution and impacting scalability, as shown in Figure~\ref{fig:granularity}(b).

Particularly noteworthy is the potential for circular dependencies to form between coarse-grained scheduling units, a phenomenon depicted in Figure~\ref{fig:granularity}. To tackle this, \system combines such dependencies into a single scheduling unit, first processing operations without dependencies. While this strategy successfully circumvents the issue of circularity, it may restrict parallelism. As such, the presence of cyclic dependencies is a key factor in \system's decision between utilizing \fsu or \csu.

\subcompact
\subsection{Abort Handling Mechanisms}
\label{subsec:abort_handling}
In \system, transaction aborts are handled in one of two ways: ``eager'' or ``lazy'' aborting.

\label{subsubsec:eager}
\label{symbol:ea}
In the eager approach (\ea), threads abort transactions as soon as an operation fails, minimizing disruption to other ongoing operations. The implementation of \ea varies based on the exploration strategy employed. Under structured exploration, \ea operates in a layered fashion. When all operations within a stratum have been executed, threads start aborting processes should an operation fail. This includes aborting the failed operation and its logical dependents and updating affected operations by rolling back and restarting from the outermost stratum with aborted operations. In contrast, under non-structured exploration, \ea employs a coordinator-based mechanism. The first operation of each state transaction is designated as the ``head'', and the thread that processes it acts as the transaction's coordinator. Should a thread detect a failed operation, it notifies the coordinator, who subsequently aborts all operations in the transaction and instructs other threads to roll back and redo dependent operations.

\label{symbol:la}
Alternatively, in the lazy approach (\la), threads log failed operations without immediately processing them. Only once the entire \tpg has been fully explored do threads collaborate to abort failed operations and their logical dependents. Although this approach is straightforward to implement and reduces context-switching overhead, it may necessitate repeated checks for transaction aborts and repeated iterations, which can lead to significant computational waste compared to \ea.

\label{subsec:rollback}
To ensure a correct schedule, irrespective of the abort handling approach, states modified by aborted operations need to be rolled back. \system accomplishes this by generating a physical copy of each modified state, with each version timestamped based on the modifying operation. Upon an operation's abort, the associated state is rolled back to the version with the latest timestamp prior to the aborted operation. These physical copies can be discarded after the current transaction batch has been completely processed. This multi-versioning state storage approach also lends itself to supporting windowing queries within \system.

\begin{table}[t]
\centering
\caption{Workload characteristics to \tpg Properties}
\label{tab:properties}
\resizebox{0.48\textwidth}{!}{
\begin{tabular}{|c|l|l|}
\hline
\textbf{Type} & {\textbf{\tpg Prop.}} & \textbf{Workload Char.} \\ \hline
\multirow{3}{*}{Vertex} & Computation Complexity & $C$ \\ \cline{2-3} 
 & Vertex Degree Distribution & $\underset{\sim}{\propto} \theta$ \\ \cline{2-3}
 & Ratio of Aborting Vertexes & $\underset{\sim}{\propto} \alpha$ \\ \hline
\multirow{4}{*}{Edge} 
 & Number of \lds & $\underset{\sim}{\propto} T*l$ \\ \cline{2-3} 
 & Number of \tds & $\underset{\sim}{\propto} T*l$ \\ \cline{2-3} 
 & Number of \pds & $\underset{\sim}{\propto} T*l*r$ \\ \cline{2-3} 
 & Cyclic Dependency & Correlated to $\theta$, $T$, $l$, $r$ \\ \hline
\end{tabular}
}
\end{table}

\begin{figure*}[t]
\centering
\begin{minipage}{\textwidth}
\includegraphics[width=\textwidth]{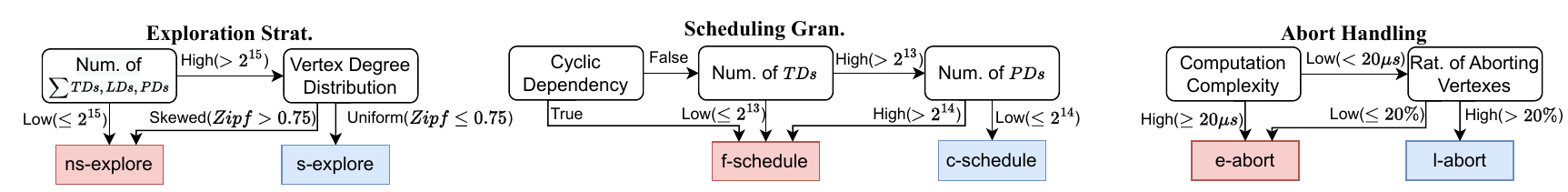}
\caption{The lightweight decision model. The concrete threshold numbers in brackets are based on our experiments.}
\label{fig:model}
\end{minipage}
\end{figure*}

\subcompact
\subsection{Heuristics Decision Model}
\label{subsec:model}
Given the NP-complete nature of the original task graph scheduling problem and the added complexity of our context, our proposed solution is a lightweight, heuristic-based decision model. Informed by comprehensive microbenchmark studies and theoretical analysis (Table~\ref{tab:decisions}), it effectively guides the \system in making scheduling decisions.

\textbf{Model Inputs.} The model operates on seven properties from the constructed \tpg (Table~\ref{tab:properties}), reflecting different workload characteristics. Vertex Computational Complexity corresponds to the complexity of the user-defined function in the associated state access operation ($C$). Vertex Degree Distribution reflects state access distribution of the operation ($\theta$), implying some states are more frequently accessed. The Ratio of Aborting Vertexes maps to the ratio of operations that need aborting ($a$), requiring profiling and estimation. The Number of \lds, \tds, \pds relates to the number of transactions arriving during the batch interval ($T$) and transaction length ($l$), with \pds also influenced by state accesses per operation ($r$). Lastly, Cyclic Dependency represents the presence of cycles when \csu is adopted, influenced by $\theta$, $T$, $l$, and $r$.

\textbf{Decision Model.} As per Figure~\ref{fig:model}, the model operates on three parallel dimensions at runtime.
\emph{I) Exploration Strategies:} \se is chosen when the dependency types are high in number and vertex degree distribution ensures thread-level workload balance. In other cases, \nse is chosen for more flexible dependency resolution.
\emph{II) Scheduling Unit Granularities:} The model selects \csu when there are no cyclic dependencies among operations, \tds number is high, and \pds number is low. For other cases, \fsu is chosen for its better scalability.
\emph{III) Abort Handling Mechanisms:} \la is chosen when computational complexity of vertices is low and the ratio of aborting vertices is high. Otherwise, \ea is chosen for its minimal impact on other operations' execution.

\compact
\section{Execution: Fine-grained Task Management}
\label{sec:execution}
\system employs sophisticated mechanisms, such as Finite State Machine (FSM) Annotations and Multi-versioning State Table Management, to ensure correct scheduling and efficient execution of operations.

\subcompact
\subsection{Finite State Machine Annotations}

\begin{table}
\centering
\caption{State Definition in the \stpg.}
\label{tab:state}
\resizebox{0.4\textwidth}{!}{%
\begin{tabular}{|p{2.4cm}|p{5cm}|}
\hline
\textbf{State} & \textbf{Definition}               \\ \hline
Blocked (\blk)                   & Operation is not ready to schedule  \\ \hline
Ready (\rdy)                     & Operation is ready to schedule               \\ \hline

Executed (\exe)                  & Operation is successfully processed                     \\ \hline
Aborted (\abt)                   & Operation is aborted  \\ \hline
\end{tabular}%
}
\end{table}

\begin{figure}[t]
	\centering
	\begin{minipage}{0.4\textwidth}
	\subfloat[Regular execution.]{
			\includegraphics[width=0.45\textwidth]
			{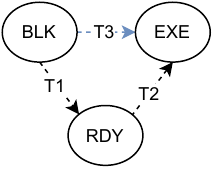}  
			\label{fig:normal}
		}	
 		\hfill
	    \subfloat[Abort and rollback.]{   
			\includegraphics[width=0.45\textwidth]
			{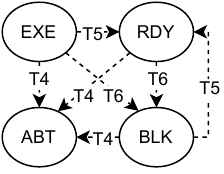}   
			\label{fig:rollback}
		}  
	\end{minipage}
	\caption{
       Six cases of state transition flow of an operation.
    }
    \label{figures:state}
\end{figure}

\system uses FSM Annotations to augment the \tpg and form an enhanced \emph{stateful task precedence graph} (\stpg). This augmentation allows \system to continuously monitor state transitions in every vertex of the \stpg, enabling the system to adjust its decisions dynamically based on current workload characteristics and maintain accurate scheduling.

Each vertex in the \stpg can be in one of four states, as summarized in Table~\ref{tab:state}:
(1) \textbf{Blocked (\blk)}, where a vertex is not ready for scheduling due to unresolved dependencies.
(2) \textbf{Ready (\rdy)}, indicating a vertex is ready for scheduling as all its dependencies have been resolved.
(3) \textbf{Executed (\exe)}, denoting a vertex has been successfully processed.
(4) \textbf{Aborted (\abt)}, representing a vertex that has been aborted due to the failed processing of itself or its dependent vertices.

\subcompact
\subsection{Multi-versioning State Table Management}
The \myc{state table management strategy} of \system incorporates multi-versioning in the state table. Each state access operation adds a new version of a record to the state table, which is annotated with a timestamp for identification. This approach allows for precise tracking and accessibility of each operation's impact on the state, offering a comprehensive history of state transitions.

This multi-versioning state table plays a crucial role in managing execution and aborting operations. When an operation transitions to the \exe state, the corresponding state in the state table is updated with a new record. Conversely, when an operation transitions to the \abt state, the subsequent records are removed from the state table.

\subcompact
\subsection{Execution Details}

\myc{
Figure~\ref{figures:state} outlines two scenarios of state transition, including six cases (\textbf{T1}$\sim$\textbf{T6}). Transitions \textbf{T1}$\sim$\textbf{T3} occur during execution, while \textbf{T4}$\sim$\textbf{T6} occur during abort handling. 
We introduce the detailed state transition with the associated state table management strategy that enables the correct execution of state transactions in this section.
}


\subsubsection{Regular Execution}
State transition during regular execution, as depicted in Figure~\ref{fig:normal}, can be categorized into two primary scheduling methods used in \system, namely, Serial and Speculative scheduling. Both techniques are coordinated under the management of \system's multi-versioning state table, which ensures the correctness of state access during the execution of state operations.

In \textbf{Serial Scheduling}, operation transitions from a \blk to \rdy state (\textbf{T1}) when all its dependent operations are in the \exe state, thereby making it ready for scheduling. Once in the \rdy state, the operation is scheduled for execution by retrieving the target states from the multi-versioning state table and applying user-defined functions among them. If successful, this processing leads to the operation's transition to the \exe state (\textbf{T2}).

\textbf{Speculative Scheduling} offers an enhancement to execution concurrency. Specifically, an operation in the \blk state can be speculatively scheduled, despite having unresolved dependencies (\textbf{T3}). If the targeted states are unavailable, the state table management strategy in \system ensures that the operation can wait until its targeted versions of states become available, although this may introduce additional context-switching overhead.

\subsubsection{Abort and Rollback}


The state transition during abort handling is illustrated in Figure~\ref{fig:rollback}.
When the processing of an operation fail or its logically dependent operations transit to \abt, the state of the operation transits to \abt from any state (\textbf{T4}). During this process, \myc{by relying on state table management strategy}, \system clears the state access impact of write operations on the state table, removing all versions of states appended after the states appended by the aborted operations.

Rollbacks, on the other hand, represent a strategic aspect of \system's abort handling. Instead of simply rolling back all affected operations to the \blk state, the system proactively checks whether the operation is ready to execute immediately after rollback, which could mean transitioning from \exe or \blk to \rdy (\textbf{T5}). However, if the dependent operations roll back to \rdy or \blk, the current operation must roll back to \blk due to unresolved dependencies (\textbf{T6}).

\subcompact
\subsection{Running Example}
\begin{figure}
\centering
\includegraphics[width=0.49\textwidth]{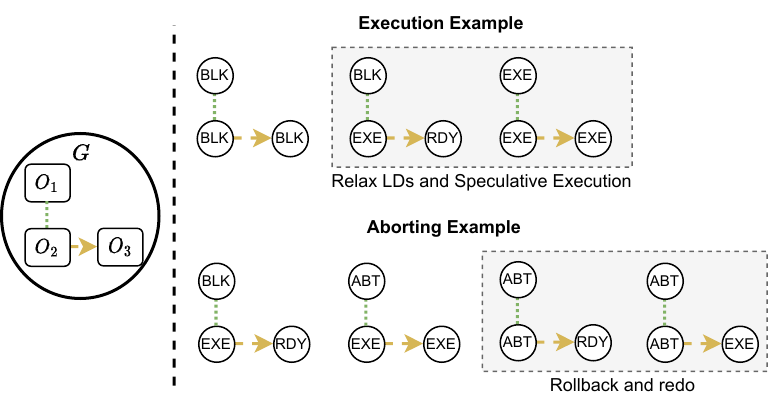}
\caption{A running example demonstrating both execution and aborting cases of state transition of operations in an \stpg.}
\label{fig:state_transition_example}
\end{figure}

To demonstrate the state transitions of operations in an \stpg, consider an example shown in Figure~\ref{fig:state_transition_example}. This example features three operations - $O_1, O_2, O_3$. Here, $O_1$ and $O_2$ are logically dependent on each other, while $O_3$ is parametrically dependent on $O_2$.

Initially, all operations are in the \blk state. Upon the first round of exploration, $O_1$ and $O_2$ transition to \rdy, as there are no \td and \pd dependencies among them. They can therefore be speculatively executed concurrently. During execution, $O_2$ transitions to \exe before $O_1$, and subsequently, $O_3$ transitions to \rdy. Once $O_1$ and $O_3$ execute and transition to \exe, the execution process based on \stpg is complete.

However, during the execution process, there may be a need for aborting and rolling back operations. For instance, when $O_1$ and $O_3$ are executed concurrently, $O_1$ may abort while $O_3$ successfully executes. Given the \ld between $O_1$ and $O_2$, $O_2$ transitions to \abt, as mandated by the aborting mechanism. Consequently, $O_3$ must rollback from \exe to \rdy due to its \pd edge with $O_2$. Finally, \system initiates a redo from $O_3$, transitioning its state to \exe and marking the completion of the execution based on \stpg.

\subcompact
\subsection{Special Scenarios}

\myc{
We then introduce the execution details of special scenarios, i.e., window and non-deterministic state operations.
}

\subsubsection{Execution of Window Operations}
Window operations are a unique execution scenario facilitated by \system's multi-versioning state table. These operations can access the desired versions of states followed by a user-defined aggregation function. Stream window queries, such as sliding and tumbling window queries, are executed by periodically triggering and processing window operations, with the triggering period depending on window size and slide size configurations. For each window operation, the targeting range of states to access and the function to be applied are specified. This process is facilitated by querying multiple versions of states on the state table.

\subsubsection{Execution of Non-deterministic State Operations}
Non-deterministic state operations also present a distinct execution scenario that necessitates additional system-level support. When executing such an operation, \system computes a key corresponding to the current moment based on workload requirements. The state accessed by non-deterministic state operations is then recorded in the \stpg, which ensures deterministic rollback in case of transaction aborts. During the state transition triggered by aborting or rolling back, \system removes the associated versions of these records to ensure correctness and consistency.

\compact
\section{Implementation Details}
\label{sec:implement}
This section presents the implementation aspects of \system. We discuss the programming model and APIs, followed by a detailed overview of the system architecture.

\begin{table}[]
\centering
\caption{User-implemented APIs}
\label{tab:user_api}
\resizebox{0.48\textwidth}{!}{%
\begin{tabular}{|p{4cm}|p{6cm}|}
\hline
APIs & Description \\ \hline

\textbf{PRE\_PROCESS} (\textcolor{blue}{Event} $e$)
& Implements a pre-processing function (e.g., filtering). Returns an EventBlotter containing parameter values (e.g., read/write sets) extracted from $e$. \\ \hline

\textbf{STATE\_ACCESS} (\textcolor{blue}{EventBlotter} $eb$)
& Facilitates state transactions by constructing system-provided APIs, such as $READ$ and $WRITE$. \\ \hline

\textbf{POST\_PROCESS} (\textcolor{blue}{Event} $e$, \textcolor{blue}{EventBlotter} $eb$)
& Implements a post-processing function that depends on the results of state access stored in the EventBlotter. \\ \hline

\end{tabular}%
}
\end{table}

\begin{table}[]
\centering
\caption{System-provided APIs (parameters for $table$, $timestamp$, and $EventBlotter$ are omitted)}
\label{tab:system_api}
\resizebox{0.48\textwidth}{!}{%
\begin{tabular}{|p{3cm}|p{5.5cm}|}
\hline
APIs & Description \\ \hline
\textbf{READ} (\textcolor{blue}{Key} $d$, \textcolor{blue}{EventBlotter} $eb$)  
&  Initiates a read request for the key $d$ and stores the results in $eb$ for further processing (i.e., post-process). \\ \hline

\textbf{WRITE} (\textcolor{blue}{Key} $d$, \textcolor{blue}{Fun} $f^*$ (\textcolor{blue}{Keys} $s...n$))
& Initiates a write request so that $state(d)$ is updated with the results after applying $f^*$ on $state(s...n)$, representing a data dependency.  \\ \hline

\textbf{READ} (\textcolor{blue}{Window\_Fun} $win\_f^*$ (\textcolor{blue}{Key} $d$, \textcolor{blue}{Size} $t$), \textcolor{blue}{EventBlotter} $eb$)
& Issues a window read request that applies $win\_f^*$ on a key $d$ with size $t$ and stores the results in $eb$ for further processing (i.e., post-process). \\ \hline

\textbf{WRITE} (\textcolor{blue}{Key} $d$, \textcolor{blue}{Window\_Fun} $win\_f^*$ (\textcolor{blue}{Keys} $s...n$, \textcolor{blue}{Size} $t$))
&  Initiates a window write request so that $state(d)$ is updated with the results after applying $win\_f^*$ on $state(s...n)$ with size $t$, this request implies data dependency. \\ \hline

\textbf{READ} (\textcolor{blue}{Fun} $f^*$, \textcolor{blue}{EventBlotter} $eb$)  
& Issues a non-deterministic read request on a key determined by a user-defined function $f^*$ and stores result in $eb$ for further processing (i.e., post-process). \\ \hline

\textbf{WRITE} (\textcolor{blue}{Fun} $f1^*$, \textcolor{blue}{Fun} $f2^*$)  
& Initiates a non-deterministic write request where the key to be updated is determined by a user-defined function $f1^*$ and the value to be written back is determined by another user-defined function $f2^*$. \\ \hline

\end{tabular}%
}
\end{table}

\subcompact
\subsection{Programming Model and APIs}
\begin{algorithm}[t]
\footnotesize
  boolean $dualmode$;\tcp{\footnotesize{\color{Gray}flag of dual-mode scheduling}}
  Map $cache$;\tcp{\footnotesize{\color{Gray}thread-local storage}}
  \ForEach{\text{event $e$ in input stream}}
  {
    \uIf{$e$ is not punctuation\tcp{\footnotesize{\color{Gray}always true under prior schemes}}}{
        EventBlotter $eb$ $\gets$ \textbf{PRE\_PROCESS}($e$);\tcp{\footnotesize{\color{Gray} e.g., filter events}}
        \textbf{STATE\_ACCESS}($eb$);\tcp{\footnotesize{\color{Gray}issue one state transaction}}
        \uIf{dualmode}{
            \tcc{\footnotesize{\color{Gray}stores events whose state access is postponed under \system scheme.}}
      	    $cache$.add($<e$, $eb>$)\;
            
      	}\Else{
      	    \tcc{\footnotesize{\color{Gray}evaluates three steps contiguously under prior schemes.}}
      	    \textbf{POST\_PROCESS}($<e$, $eb>$);\tcp{\footnotesize{\color{Gray}e.g., computes toll based on obtained road statistics}}
      	}
    }
    \Else{
        \tcc{\color{Gray}if the event is a punctuation, transaction processing can start.}
        \textbf{TXN\_START}()\tcp{\color{Gray}Triggers mode switching.}
        \ForEach{$<e$, $eb>$ $\in$ $cache$}{
            \textbf{POST\_PROCESS}($<e$, $eb>$);
        }
    }
  }
  \caption{Code template of an operator}
  \label{alg:algo_example}
\end{algorithm}

\begin{algorithm}[t]
\footnotesize
    \KwIn{EventBlotter $eb$}
    \SetKwFunction{FMain}{Sender\_Fun}
    \SetKwProg{Fn}{Function}{:}{}
    \Fn{\FMain{$sender$, $value$}}{
        \uIf{\textbf{READ}($sender$, $eb$) $>$ $value$}{
            return \textbf{READ}($sender$, $eb$) - $value$\;\tcp{\footnotesize{\color{Gray}decrement money of $sender$ by $value$.}}
        }
    }
   
    \SetKwFunction{FMain}{Recver\_Fun}
    \SetKwProg{Fn}{Function}{:}{}
    \Fn{\FMain{$recver$, $sender$, $value$}}{
        \uIf{\textbf{READ}($sender$, $eb$) $>$ $value$}{
            return \textbf{READ}($recver$, $eb$) + $value$\;\tcp{\footnotesize{\color{Gray}increment money of $recver$ by $value$.}}
        }
    }

    \Begin{
         \textbf{WRITE}($eb.sender$, Sender\_Fun($eb.sender$, $eb.v$))\;
         \textbf{WRITE}($eb.recver$, Recver\_Fun($eb.recver$, $eb.sender$, $eb.v$))\;
     }
 \caption{\footnotesize{STATE\_ACCESS of \texttt{\myc{Streaming Ledger}}}}    
  \label{alg:algo2}
\end{algorithm}

\begin{algorithm}[t]
\footnotesize
    \KwIn{EventBlotter $eb$}

    \SetKwFunction{FMain}{Fun1}
    \SetKwProg{Fn}{Function}{:}{}
    \Fn{\FMain{$udf1$}}{
         $target\_key$ $\gets$ $udf1$\;\tcp{\footnotesize{\color{Gray}get the targeting key to write through $udf1$.}}
         return $target\_key$\;
    }

    \SetKwFunction{FMain}{Fun2}
    \SetKwProg{Fn}{Function}{:}{}
    \Fn{\FMain{$udf2$}}{
         $keys$ $\gets$ $udf2$\;\tcp{\footnotesize{\color{Gray}get keys to read through $udf2$.}}
         $sum\_value$ $\gets$ \textbf{sum} \textbf{READ}($keys$)\;
         $eb.result$.insert($sum\_value$)\;
         return $sum\_value$\;
    }

    \Begin{
        \textbf{WRITE}(Fun1($eb.udf1$), Fun2($eb.udf2$));
     }
 \caption{\footnotesize{STATE\_ACCESS of \texttt{Non-deterministic GrepSum}}}    
  \label{alg:algo3}
\end{algorithm}

\system, similar to many popular DSPEs~\cite{storm,carbone2015apache,spark}, expresses an application as a Directed Acyclic Graph (DAG), where vertices are operators encapsulating user-defined data processing logic, and edges are the event streams connecting these operators. Inside each operator, we encapsulate the data processing logic into a three-step programming model. These steps include preprocessing, state access, and postprocessing, which are recursively applied to every batch of input events. 

In the \textit{preprocessing} step, \system identifies the potential read/write sets from the input events following the predefined event parsing logic. It is noteworthy that these sets may contain non-deterministic functions and thus their concrete values may not be fully determined during preprocessing. 
The second step, \textit{state access}, is where actual state accesses occur. Here, \system applies user-defined functions to each identified read/write set from the preprocessing stage. This state access step is comprehensive and capable of handling special operations such as non-deterministic and window operations. Non-deterministic state accesses, identified during preprocessing, are resolved in this step to achieve concrete state accesses. 
\textit{Postprocessing} is the final step. Here, \system further processes the input event according to the results from state access, generating the corresponding output. If a transaction aborts, \system marks the output with ``failed state access'' to notify users, who can resubmit requests as part of new input events.

\system enhances this programming model by providing a list of APIs, both user-implemented and system-provided. User-implemented APIs, outlined in Table~\ref{tab:user_api}, demand users to incorporate their application requirements into the three-step data processing logic. Note that, \system maintains a thread-local auxiliary data structure called \emph{EventBlotter} for each input event, which acts as the data bridge linking the two processing phases as discussed in Section~\ref{subsec:overview}. An example of an operator's code template based on this programming model is shown in Algorithm~\ref{alg:algo_example}. State transaction execution is conveyed through the \textbf{STATE\_ACCESS} API, which is further defined using system-provided APIs. 
\myc{Algorithm~\ref{alg:algo2} and~\ref{alg:algo3} provide an illustration of \textbf{STATE\_ACCESS} implementation using two examples, namely the \texttt{Streaming Ledger} and the \texttt{Non-deterministic GrepSum}. The \texttt{Non-deterministic GrepSum} is a modified version with non-deterministic operations atop the original GrepSum benchmark introduced in our previous work~\cite{tstream}.}

In contrast, system-provided APIs, summarized in Table~\ref{tab:system_api}, stand as library calls, offering functions akin to those in other frameworks~\cite{he2008mars}. \textbf{READ} and \textbf{WRITE} represent atomic operations that a state transaction can decompose. \system's APIs go beyond normal \textbf{READ}/\textbf{WRITE} operations by supporting window and non-deterministic operations, extending the flexibility and adaptability of the platform. For window operations, users can define a window function in \textbf{READ}/\textbf{WRITE} operations, specifying states to access, window size, and a window aggregation function. Non-deterministic operations allow users to specify user-defined functions for both keys to access and values to write back. This indicates that while the number of operations a transaction decomposes is deterministic according to the processing logic, the accessed keys and aggregated results of operations can be non-deterministic.

\subcompact
\subsection{System Architecture}
\label{subsec:architecture_overview}
The architecture of  \system is composed of various components working in unison to support the efficient execution of transactional stream processing. This section provides a detailed overview of the key architectural components and outlines their interactions and workflow in \system.

\subsubsection{Architectural Components and Interactions}

\begin{figure*}[t]
\centering
\includegraphics[width=0.9\textwidth]{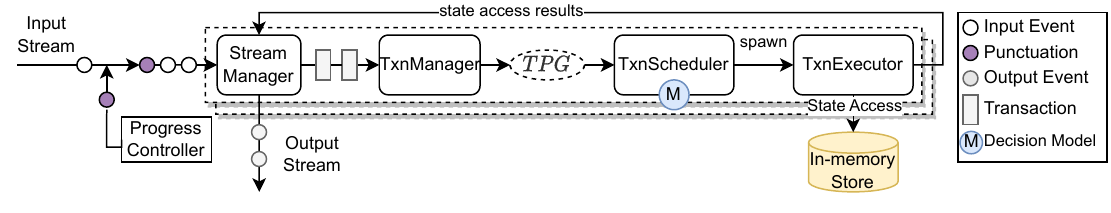}
\caption{The system architecture of \system. The constructed \tpg and shared state are stored in memory.}
\label{fig:architecture_overview}
\end{figure*}

\myc{An overview of the system architecture of \system can be shown in Figure~\ref{fig:architecture_overview}}, which is comprised of five primary components: the singleton ProgressController (PC), and per-thread instances of StreamManager (SM), TxnManager (TM), TxnScheduler (TS), and TxnExecutor (TE).

The PC has a pivotal role, where it assigns monotonically increasing timestamps to its generated punctuations using a simple global counter. Its operation ensures a coherent temporal view across all threads in the system.

SM is the first stage in the transactional processing pipeline. It handles the preprocessing and postprocessing of each input event ($e$). During preprocessing, SM can generate state transactions, an approach adapted from prior works~\cite{tstream}. However, unlike traditional models, these transactions are not immediately processed. Instead, the input events are postprocessed based on the results of these state transactions once they are processed, i.e., either committed or aborted.

The responsibility of dependency resolution among state transactions and construction of a \tpg falls on TM. TM's operation comes into play upon receiving punctuation, during which it refines the constructed \tpg with additional dependency resolution. This refined \tpg serves as the basis for the next stage of the pipeline, the TS, which makes scheduling decisions based on the \tpg, guiding the system towards concurrent operation execution, a feature further discussed in Section~\ref{sec:scheduling}. 

Rounding off the pipeline is TE, spawned by TS, which executes operations in \tpg according to the scheduling decisions. It ensures correct state access/abort for operations in \tpg with finite state machine annotations. The details of this process are presented in Section~\ref{sec:execution}.

\begin{algorithm}[t]
\footnotesize
    \KwData{$e$ \tcp{Input event}}
    \KwData{$txn_{ts}$ \tcp{State transaction}}
    \KwData{$G$ \tcp{The currently constructed \tpg}}
    \While{!finish processing of input streams}{
        \eIf(\tcp*[h]{Phase 1}){\text{$e$ is not a $punctuation$}}{
                $txn_{ts}$ $\gets$ PRE\_Processing($e$)\;
                \tcp*[h]{Planning: dependency tracking}\;
                \textbf{\circled{1}$^1$}\textbf{TPG\_Construction}($G$, $txn_{ts}$)\; 
          }(\tcp*[h]{Phase 2}){
                \textbf{\circled{1}$^2$}\textbf{TPG\_Refinement}($G$)\; 
                \tcp*[h]{Scheduling: adaptive scheduling}\;
                \textbf{\circled{2}}$M$ $\gets$ \textbf{TPG\_Scheduling}($G$)\;
                \tcp*[h]{Execution: correct execution}\;
                \textbf{\circled{3}}\textbf{TPG\_Execution}($G$, $M$)\; 
                POST\_Processing()\;
          }
    }
    
    \SetKwFunction{FMain}{TPG\_Construction}
    \SetKwProg{Fn}{Function}{:}{}
    \Fn{\FMain{$G$, $txn_{ts}$}}{
        $O_{1..k}$ $\gets$ \textbf{Partition} $txn_{ts}$\;
        \ForEach{\text{operation $O_{i}$ $\in$ $O_{1..k}$}}{
            \textbf{Identify} its \ld\;
            $G$ $\gets$ $G$ + $O_{i}$\;
        }
    }
    \SetKwFunction{FMain}{TPG\_Refinement}
    \SetKwProg{Fn}{Function}{:}{}
    \Fn{\FMain{$G$}}{
        \ForEach{\text{vertex $e_{i}$ $\in$ $G$}}{
            \textbf{Identify} its \td, \pd\;
        }
    }

    \SetKwFunction{FMain}{TPG\_Scheduling}
    \SetKwProg{Fn}{Function}{:}{}
    \Fn{\FMain{$G$}}{
        $M$ $\gets$ Instantiated with $G$;\tcp{Model-guided scheduling decision making.}
    }
    
    \SetKwFunction{FMain}{TPG\_Execution}
    \SetKwProg{Fn}{Function}{:}{}
    \Fn{\FMain{$G$, $M$}}{
        \While{!finish scheduling of $G$
        }{
        $Scheduling Unit$ $\gets$ \emph{Explore}($G$, $M$)\;
        Execute with abort handling ($Scheduling Unit$)\; 
        }
    }
  \caption{Execution Algorithm of \system}
  \label{alg:algo}
\end{algorithm}

\subsubsection{Execution Algorithm}
\label{subsec:execution_algorithm}
In this section, we detail the execution algorithm for \system, which drives the operation of the system's components in line with the three-stages execution workflow (as detailed in Section~\ref{subsec:overview}). This algorithm plays a pivotal role in ensuring the correct and efficient processing of transactions within the stream processing context. Each phase in the workflow — Planning, Scheduling, and Execution — corresponds to a specific set of tasks performed by the algorithm, which leverages the system components introduced above and handles the dependencies among state transactions as identified in the \tpg. In the following paragraphs, we elucidate the algorithm, with particular emphasis on its connection to the three-stage workflow.

The primary execution algorithm of \system, as outlined in Algorithm~\ref{alg:algo}, underscores that transactional stream processing is a continuous process that may potentially never end. By introducing punctuations~\cite{Tucker:2003:EPS:776752.776780}, the algorithm separates the dependency resolution and execution of state transactions into two non-overlapping phases, ensuring that no subsequent input event will have a smaller timestamp.

In the first phase, the \textbf{\circled{1}}$^1$ \textit{StreamManager} preprocesses every input event ($e$). Concurrently, the \textbf{\circled{1}}$^1$ \textit{TxnManager} manages dependency resolution among state transactions and integrates decomposed operations to construct a \tpg.

The second phase sees the \textbf{\circled{1}}$^2$ \textit{TxnManager} refining the constructed \tpg through further dependency resolution. Meanwhile, the \textbf{\circled{2}} \textit{TxnScheduler} schedules operations for concurrent execution based on the refined \tpg. 

In \textbf{\circled{3}}, guided by a scheduling decision model $M$, execution threads adopt an exploration strategy to traverse the constructed \tpg, identifying operations ready to be scheduled while respecting dependency constraints. During this exploration, one or several operations may be treated as the unit of scheduling. Each thread executes the operations in the scheduling unit, implementing various abort handling mechanisms as needed. Input events associated with processed state transactions (i.e., committed or aborted) are then postprocessed by the \textit{StreamManager} according to the transaction processing results.

\compact
\section{Evaluation}
\label{sec:exp}
In this section, we conduct a comprehensive evaluation of \system, comparing it to alternative approaches. Note that these experiments are conducted under the assumption of no system failures during runtime. Providing efficient fault tolerance without compromising low latency and high throughput during normal operation poses a significant challenge for TSPEs, even in a single-node setting. This complexity arises from the intricate interplay of transactional and stream-oriented properties that TSPEs exhibit. The challenge could potentially magnify in a distributed environment. Enhancements for \system's fault tolerance are a subject of our separate, forthcoming work.

In summary, we have made the following key observations.
\begin{itemize}
    \item Our experimental results show that \system outperforms conventional SPEs for TSP applications (Section~\ref{subsubsec:conventional}) by orders of magnitude. Because of the adaptive scheduling strategy, \system achieves up to 2.2x higher throughput and 69.1\% lower latency compared to the state-of-the-art TSPEs (Section~\ref{subsubsec:dynamic} and~\ref{subsubsec:mutiple}).
    \myc{Moreover, we also evaluate the performance of executing window non-deterministic queries on \system in Section~\ref{subsubsec:window_exp} and~\ref{subsubsec:nondeterministic_exp}.}
    \item In Section~\ref{subsec:overhead}, we show that \system spends more time on \tpg construction and exploration, but largely reduces the overhead of synchronization. A drawback of \system, however, is its high memory consumption (about 1.4x times higher) due to the more complex auxiliary data structures.
    \item We show that no one scheduling strategy can outperform others in all cases (Section~\ref{subsec:scheduling_decision}). Each scheduling decision has its own advantages and disadvantages under varying workload characteristics.
    \item In Section~\ref{subsec:modern_hardware}, we show that \system spends up to 2.3x fewer clock ticks and has a lower memory bound than \tstream and \sstore. Furthermore, \system has much better multicore scalability compared to prior schemes. 
    \item \myc{In Section~\ref{subsec:use_cases}, we demonstrate the practical usefulness of \system with two case studies: \textit{Online Social Event Detection} and \textit{Stock Exchange Analysis}. Specifically, \system can react to emerging events and output expected results in sub-second level when processing real-world workloads.}
\end{itemize}
 
\subcompact
\subsection{Evaluation Methodology}
We conduct all experiments on a dual-socket Intel Xeon Gold 6248R server with 384 GB DRAM. 
Each socket contains $24$ cores of 3.00GHz and 35.75MB of L3 cache. To isolate the impact of NUMA, we use one socket of the server in our experiments. We leave it as a future work to address the NUMA effect~\cite{briskstream}. We pin each thread on one core and assign 1 to 24 cores to evaluate the system scalability.
The OS kernel is \emph{Linux 4.15.0-118-generic}. 
We use \emph{JDK 1.8.0\_301}, set \emph{-Xmx} and \emph{-Xms} to be 300 GB. We use G1GC as the garbage collector across all the experiments and configure \system to not clear temporal objects such as the processed TPGs and multi-versions of states. 
We show the impact of clean-up and JVM GC in Section~\ref{subsec:memory-footprint}.

We use three use cases: Streaming Ledger (SL), GrepSum (GS), Toll Processing (TP) from a benchmark proposed by our previous work~\cite{tstream} on evaluating the effectiveness of \system. For all these workloads, we follow the original application logic but tweak the configurations to bring more workload dependencies such that we can better expose the issues of existing TSPEs. In particular, 
we have configured ten times larger sizes of shared mutable states and generated more state transactions accessing overlapped states in our workload settings.
Additionally, we further present two additional use cases that process the real-world datasets: Online Social Event Detection~\cite{sahin2019streaming,olteanu2014crisislex} (OSED) and Stock Exchange Analysis~\cite{sse} (SEA) to illustrate the usefulness of \system in supporting complex real-world data processing scenarios.

\textbf{\change{Tuning Workload Characteristics.}}
To better comprehend the system behaviour, we tune the following six workload characteristics. The default workload characteristics and varying ranges are summarized in Table~\ref{tab:default}. 
\emph{1). State Access Distribution ($\theta$):}
Similar to~\cite{tstream}, we modelled the state access distribution as Zipfian skew, and tune the Zipfian factor to vary $\theta$.
\emph{2). Ratio of Aborting Transactions ($a$):} 
We tune $a$ by artificially adding transactions that violate the consistency property, such as the account balance can not become negative.
\emph{3). Transaction Length ($l$):}
We tune $l$ by varying the number of atomic state access operations in one transaction.
\emph{4). The complexity of a UDF ($C$):}
We tune $C$ by adding a random delay in each user-defined function (i.e., the $f$ in Definition~\ref{def:PD}).
\emph{5). Number of State Access Per Operation ($r$):}
We vary the ratio of multiple state access operations to tune $r$.
\emph{6). Number of Transactions ($T$):} 
We tune $T$ by varying the punctuation interval.


\textbf{Dynamic Workload Configurations.}
To evaluate the adaptability of \system, we follow the mechanism proposed by Ding et al.~\cite{dynamicWorkload} to generate dynamic workloads.
Specifically, we generate various phases of dynamic workloads by different trends, which determines the parameters we want to tune and how they change over time. For example, in a dynamic workload with an increasing tendency to abort transactions, we will increase the ratio of aborting transactions over time. 



\begin{table}[]
\centering
\caption{Workload default configuration}
\label{tab:default}
\resizebox{0.48\textwidth}{!}{
\begin{tabular}{|p{1.5cm}|p{1.8cm}|p{1.8cm}|p{1.8cm}|p{1.8cm}|p{2cm}|}
\hline
\textbf{Workload Char.} & \textbf{SL} & \textbf{GS} & \textbf{TP}  & \textbf{Tweaking ranges} \\ \hline
$\theta$ & 0.20 & 0.20 & 0.20 & 0.0$\sim$1.0 \\ \hline
$a$ & 1\% & 1\% & 1\%  & 0$\sim$90\% \\ \hline
$l$ & 2 / 4 & 1 & 2  & 1$\sim$10 \\ \hline
$C$ & 10 us & 10 us & 10 us & 0$\sim$100 us \\ \hline
$r$ & 1 / 2  & 2 & 1  & 1$\sim$10 \\ \hline
$T$ & 10240  & 10240 & 40960  & 5120$\sim$81920 \\ \hline
\end{tabular}
}
\end{table} 


\subcompact
\subsection{Performance Evaluation}
\label{subsec:evaluation}
In this section, we conduct a series of experiments to confirm \system's superiority compared to the state-of-the-art.

\subsubsection{Comparing to Conventional SPEs}
\label{subsubsec:conventional}
\begin{figure}[t]
\centering
\includegraphics[width=0.48\textwidth]{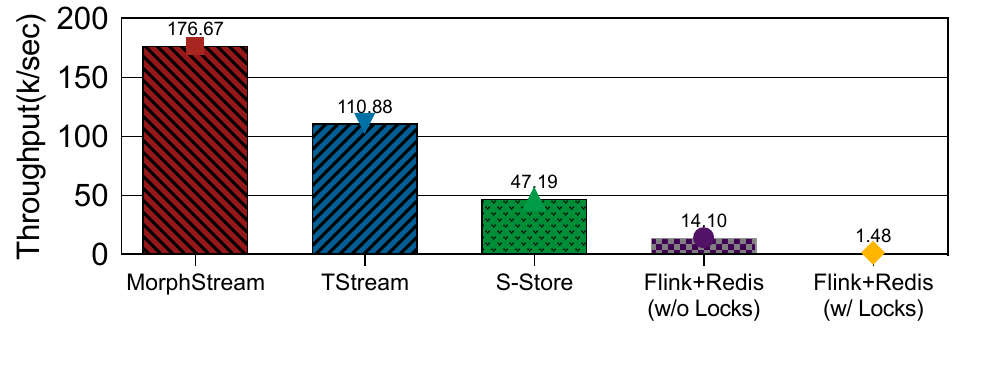}
\caption{Performance comparison among \system and existing systems for running SL on 24 cores.}
\label{fig:overview_comparison}
\end{figure}

In the first experiment, we show that TSPEs significantly outperform conventional SPEs when handling TSP applications. We use the default workload configuration shown in Table~\ref{tab:default} in this study. 
We have implemented SL on Flink-1.10.0.
Since the native Flink does not support shared mutable state accesses, we leveraged Redis-6.2.6 with a distributed lock, a common workaround, to store shared mutable states.
We deploy a standalone cluster with a single TaskManager configured with 24 slots and set the parallelism of SL to 24.
To avoid the OOM exception, we set the TaskManager heap size to 100GB.
When locking is disabled (denoted as w/o Locks), execution correctness is not guaranteed in Flink.
The detailed workload configuration is shown in Table~\ref{tab:default}.
As shown in Figure~\ref{fig:overview_comparison}, \system significantly outperforms the two state-of-the-art TSPEs, \tstream (1.6x) and \sstore (3.7x), and Flink (up to 117x).
It is noteworthy that Flink, a popular conventional SPE, achieves orders of magnitude lower throughput in this application. By disabling locks, its throughput increases but is still far lower than any of the TSPEs. In the following, we hence do not further compare \system with Flink.

\subsubsection{Evaluation on Dynamic Workloads}
\label{subsubsec:dynamic}
\begin{figure}[t]
\begin{center}
		{\setlength{\fboxsep}{1pt}
			\hspace{0.99em}
			 \vspace{-1.5em}
				\begin{minipage}{0.3\textwidth}
					\includegraphics[width=\columnwidth]
					{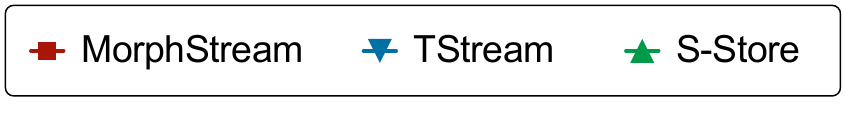}
				\end{minipage}
		}
\end{center}
\centering
	 \begin{minipage}{0.5\textwidth}
  \centering
	\subfloat[Throughput]{
		\includegraphics[width=0.81\textwidth]{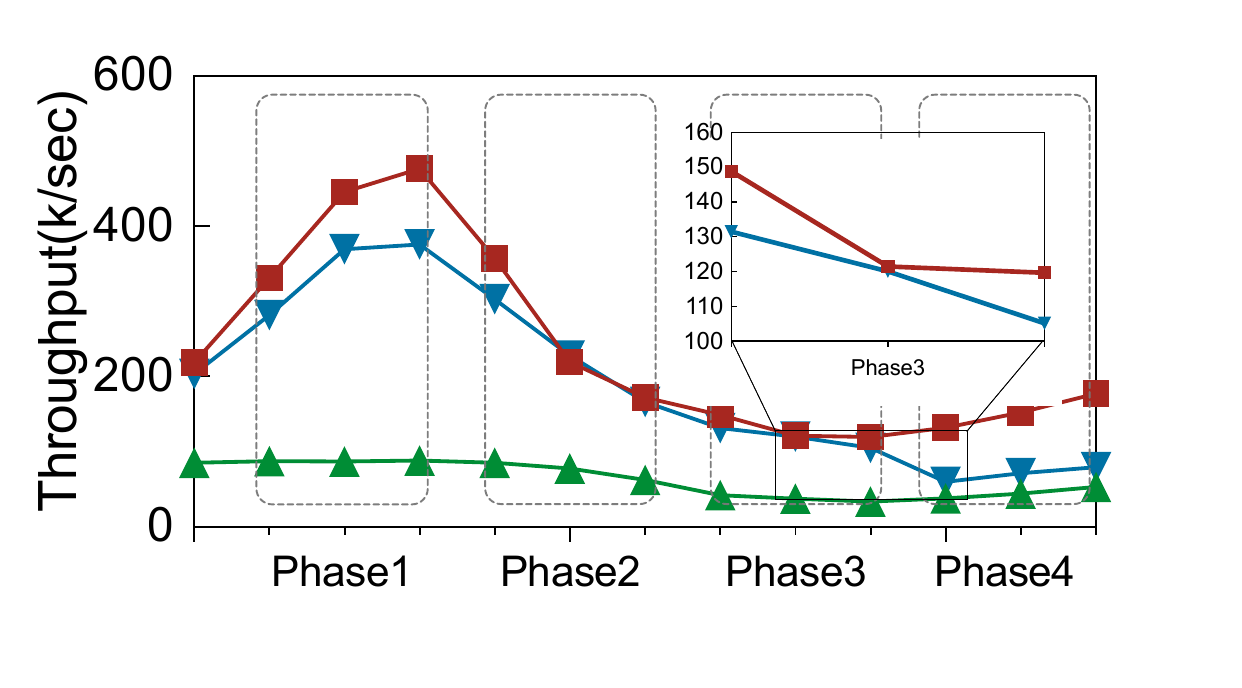}
            \label{fig:SL_Throughput}
		}	
  \end{minipage}
	 \begin{minipage}{0.5\textwidth}
  \centering
         \subfloat[Latency]{   
		\includegraphics[width=0.79\textwidth]{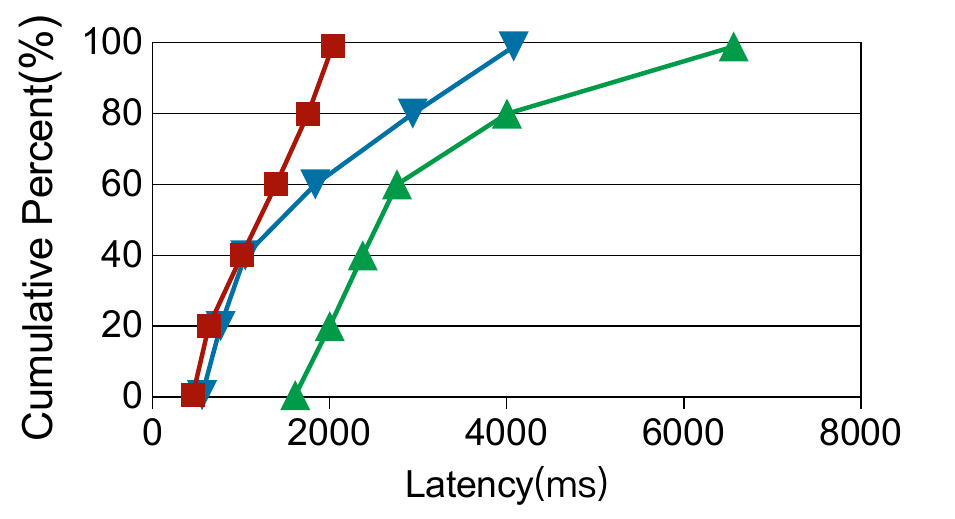}   
		\label{fig:SL_Latency}
	    }
	 \end{minipage}
	\caption{Evaluation on Dynamic Workload.}
    \label{figures:SL}
\end{figure}

In this experiment, we show that \system can always select a better-performing scheduling strategy under changing workloads, resulting in lower latency and higher throughput compared to state-of-the-art TSPEs. We use SL as the base application and divide the workloads into four phases. 
Figure \ref{fig:SL_Throughput} and Figure \ref{fig:SL_Latency} compare the throughput and latency of \system against two state-of-the-art TSPEs: \sstore~\cite{S-Store} and \tstream~\cite{tstream}. We mark each phase in the dynamic workload using the dotted grey box.
In the first phase, a large number of events consisting of \emph{Deposit} transactions arrive, and the state accesses distribution is scattered. 
As a result, there are lots of \lds and \tds but few \pds. At the same time, the vertex degree distribution is uniform as the state accesses are scattered. 
As guided by our decision model (Figure~\ref{fig:model}), 
\system selects the \se strategy to resolve a large number of dependencies and selects \csu
for scheduling since there are fewer \pds. 
\system achieves up to 1.27 times higher throughput compared to the second-best. 
In the second phase, we configure the workload with increasing key skewness over time. Hence,
dependencies are gradually contented among a small set of states, which facilitates the resolution of dependencies.
As expected, the performance of all approaches drops. 
\system gradually morphs from \se to \nse strategy to resolve dependencies in a more flexible manner, and constantly outperforms \sstore. 
In the third phase, we configure the workload with an increasing ratio of \emph{Transfer} transactions so that one of the two types of transactions in SL is called intensively in a short period of time. 
As the proportion of \emph{Transfer} transactions increases, there are more and more dependencies between scheduling units. Hence, \system gradually morphs from \csu to \fsu to reduce the dependency resolution overhead and result in a stable throughput.

There is no transaction abort in the first three phases, and the selection of aborting mechanism in \system does not matter. In the fourth phase, we increase the ratio of aborting transactions over time to evaluate the performance of the system under a dynamically changing ratio of aborting transactions. 
In the beginning, \system applies the \ea mechanism to eagerly abort when the operation fails and morphs to \la when aborts are frequent so that transaction aborts can be handled together to reduce context switching overhead.
The results show that \tstream's performance drops when transaction aborts appear. 
This is because of the rapidly increasing overhead of redoing the entire batch of transactions. 
In contrast, \system achieves relatively stable performance and is 2.2x to 3.4x higher than other schemes.
 
A further key takeaway from Figure \ref{fig:SL_Latency} is that \tstream and \sstore have significantly higher tail latency than \system. 
This is mainly because the scheduling strategies in \tstream and \sstore are limited for specific workload characteristics. When workload changes, such as increasing transaction aborts or key skewness, their efficiency drops significantly, resulting in higher processing latency. In contrast, \system dynamically morphs the scheduling strategy according to the change of workload characteristics to deal with different situations, thus achieving a constantly lower processing latency.

\begin{figure}[t]
\centering
{\setlength{\fboxsep}{1pt}
			\hspace{0.99em}
				\begin{minipage}{0.28\textwidth}
					\includegraphics[width=\columnwidth]
					{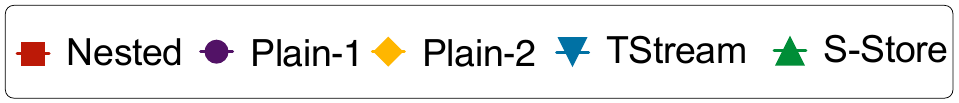}
				\end{minipage}
		}

 	\begin{minipage}[c]{0.5\textwidth}
        \centering
	\subfloat[Throughput]{
	\includegraphics[width=0.79\textwidth]{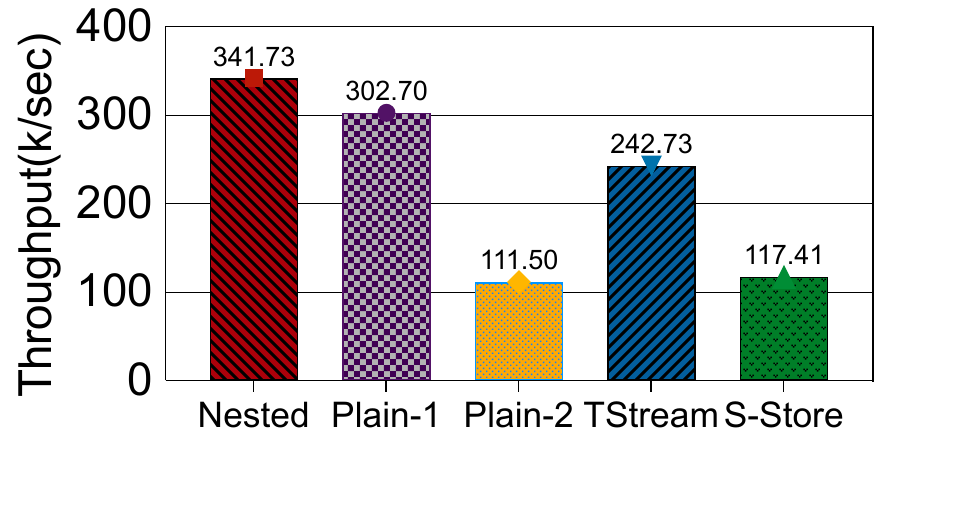}
    \label{fig:TP_Throughput}
	}	
 
    \subfloat[Latency]{   
		\includegraphics[width=0.79\textwidth]{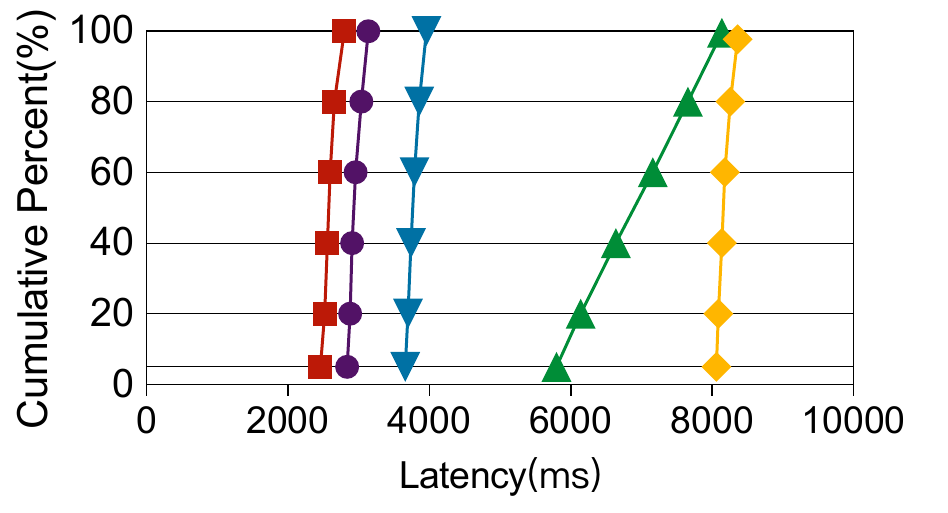}   
		\label{fig:TP_Latency}
	}
 	\end{minipage}
	\caption{Single (plain, \tstream, \sstore) vs. Multiple (nested) Scheduling Strategies.}
    \label{figures:SL}
\end{figure}
\subcompact
\subsubsection{Evaluation of Multiple Scheduling Strategies}
\label{subsubsec:mutiple}
In TP, the road conditions in different regions can have varying characteristics, which can be divided into multiple groups. \tstream \cite{tstream} and \sstore \cite{S-Store}'s scheduling strategies may work well on one group of state transactions but not on another.
In contrast, \system's modular and flexible design allows it to employ multiple scheduling strategies concurrently.
For illustration, we configure the TP to contain two groups of state transactions simultaneously. 
In \emph{group 1}, the state access distribution of state transactions is skewed, and the ratio of aborting transactions is high. 
In \emph{group 2}, the state distribution of state transactions is uniform and transaction aborts occur rarely. 
\margii{-1pt}{R4O14}
\change{As guided by our decision model (Figure~\ref{fig:model})}, 
\system applies \nse, \csu, and \la for handling transactions of \emph{group 1}, and applies \se, \csu, and \ea for handling transactions of \emph{group 2}. We name such a combination of strategies a \emph{nested} configuration.

Figure~\ref{fig:TP_Throughput} shows the throughput comparison results. We can see that the throughput of the nested configuration is 40.9\% higher than \tstream and 117\% higher than \sstore. 
To further comprehend the advantage of the nested configuration, we compare it against two plain scheduling strategies: \nse, \csu, and \la (denoted as \emph{plain-1}) and \se, \csu, and \ea (denoted as \emph{plain-2}) for handling all transactions from both groups.
Unsurprisingly, the throughput of the nested configuration is 1.17$\times$ and 2.87$\times$ higher than that of each plain scheduling strategy. When the ratio of aborting transactions in \emph{group 1} is high, the \emph{plain-2} is bottlenecked by the frequent context switching overheads. 
At the same time, as the skewness of state access increases in \emph{group 1}, the workloads become less balanced among threads, hampering the system performance when using \se in \emph{plain-2}.
As the state distribution of state transactions is uniform and the ratio of aborting transactions is low in \emph{group 2}, the \emph{plain-1} spends more time resolving dependencies and redoing the entire batch of transactions. 
The \emph{plain-1} performs better than \emph{plain-2} as there are fewer \pds and the computation complexity is low, but it is still lower than that of the \emph{nested} setting. 

Figure~\ref{fig:TP_Latency} shows the comparison results of end-to-end processing latency. Thanks to the significantly improved performance, \system with nested configuration achieves very low processing latency. 
Note that, \sstore spends more time on synchronization and inserting locks under a highly contended workload in \emph{group1} because dependent transactions are executed serially. Under a higher ratio of aborting transactions in \emph{group 1}, \emph{plain-2} spends lots of time achieving fine-grained state rollback because of the high context-switching and synchronization overhead of \se (Table~\ref{tab:decisions}), which is why \emph{plain-2} results in highest latency compared to others. 
\begin{figure}[t]
	\begin{minipage}{0.5\textwidth}
\centering
	\subfloat[Window Size]{
		\includegraphics[width=0.79\textwidth]{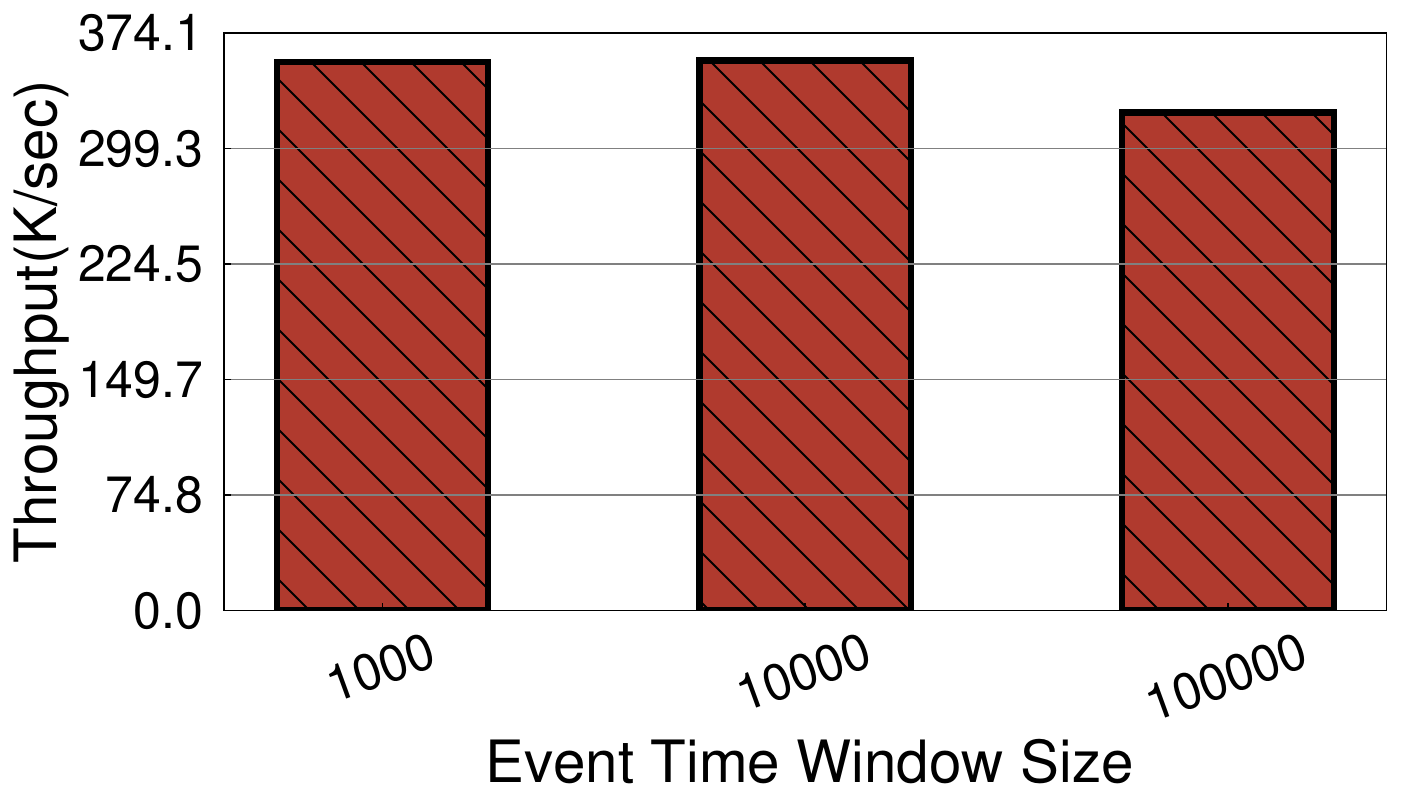}
            \label{fig:window_size_exp}}
            
        \subfloat[Window Trigger Period]{   
		\includegraphics[width=0.79\textwidth]{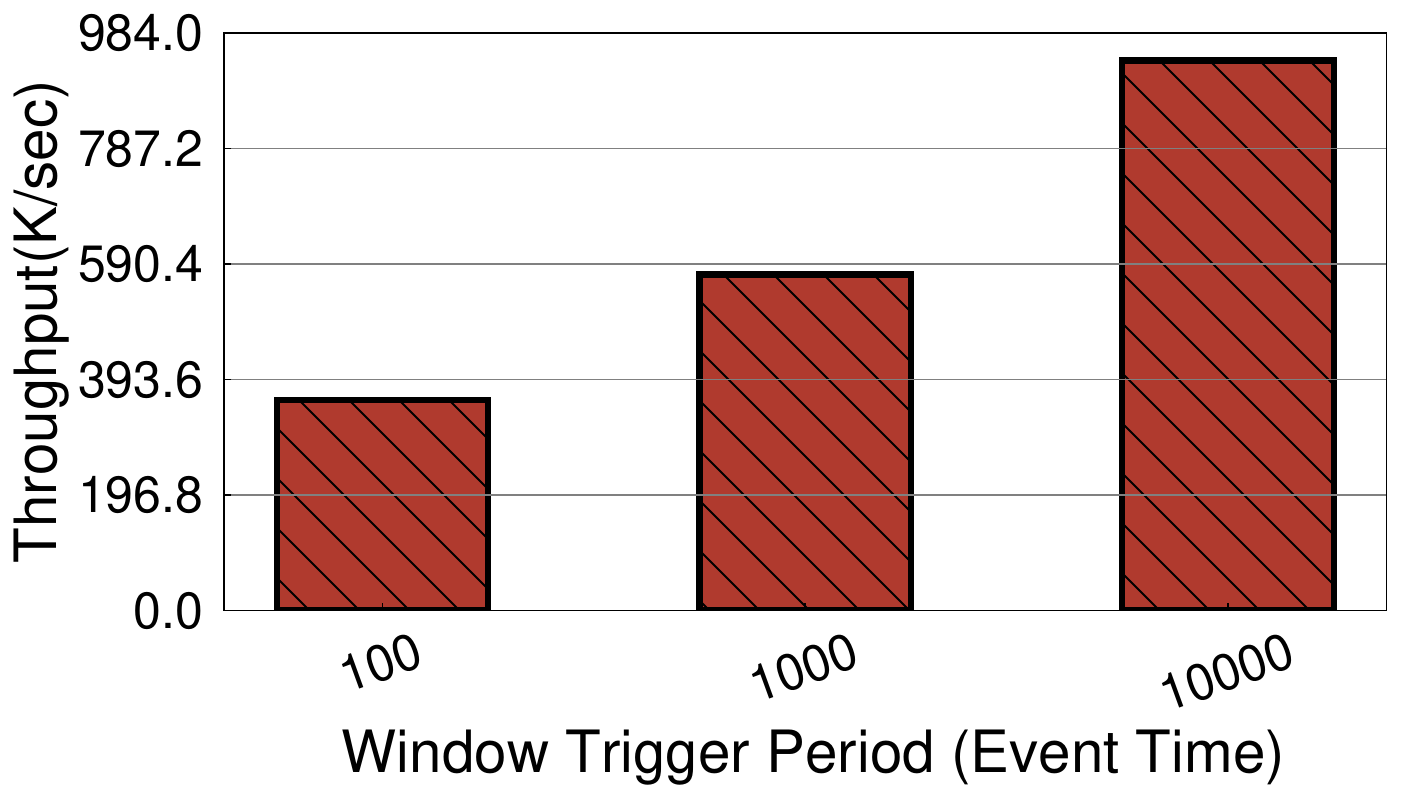}
		\label{fig:window_period_exp}}
        \caption{Evaluation of Tumbling Window Queries.}
        \label{figures:window_exp}
	\end{minipage}
\end{figure}

\subcompact
\subsubsection{Evaluation of Window-based Queries}
\label{subsubsec:window_exp}

\myc{
We implement an additional application \textit{GrepSum with window reads} as an example to illustrate how MorphStream supports windowing queries. 
Specifically, we modified the GrepSum to perform random state updates and periodical window readings. 
In particular, the application processes two types of state transactions: 
(1) it executes transactions with write-only operations on receiving input events with updating requests, 
and (2) it executes transactions with window-read and sum aggregation operations on receiving input events with
reading requests.
}

For this experiment, we retained the settings of GrepSum outlined in Table~\ref{tab:default}, with an abort ratio of 0 and a punctuation interval of 102400. The process involved periodic state access (one event with reading requests for every 100 input events), where 100 random states were accessed within a default window size of 1000 (which required reading states up to 1000 event-time old) for the GrepSum operation. We modified the window trigger periods and window sizes to simulate a variety of window query scenarios, and evaluated their impact on performance. The overall results are presented in Figure~\ref{figures:window_exp}.

We first adjust window sizes from 1k to 100k, with the performance outcomes depicted in Figure~\ref{fig:window_size_exp}. As anticipated, increasing the window size led to an escalation in state access overhead due to the need to read more state versions. This inflated overhead, in turn, decreased system throughput by up to 30\%.
We then vary the window trigger period from 100 to 10k events. The performance results, shown in Figure~\ref{fig:window_period_exp}, demonstrated that frequent window queries significantly impeded throughput, slowing it down by as much as 60\%.
These experiments underscore \system's optimization potential in window-based transactional stream processing. Opportunities lie in reducing redundant calculations in overlapping window operations. Exploring these improvements is part of our future work.

\subsubsection{Evaluation of Non-Deterministic Queries}
\label{subsubsec:nondeterministic_exp}

\begin{figure}[t]
\begin{minipage}{0.5\textwidth}
\centering
\includegraphics[width=0.69\textwidth]{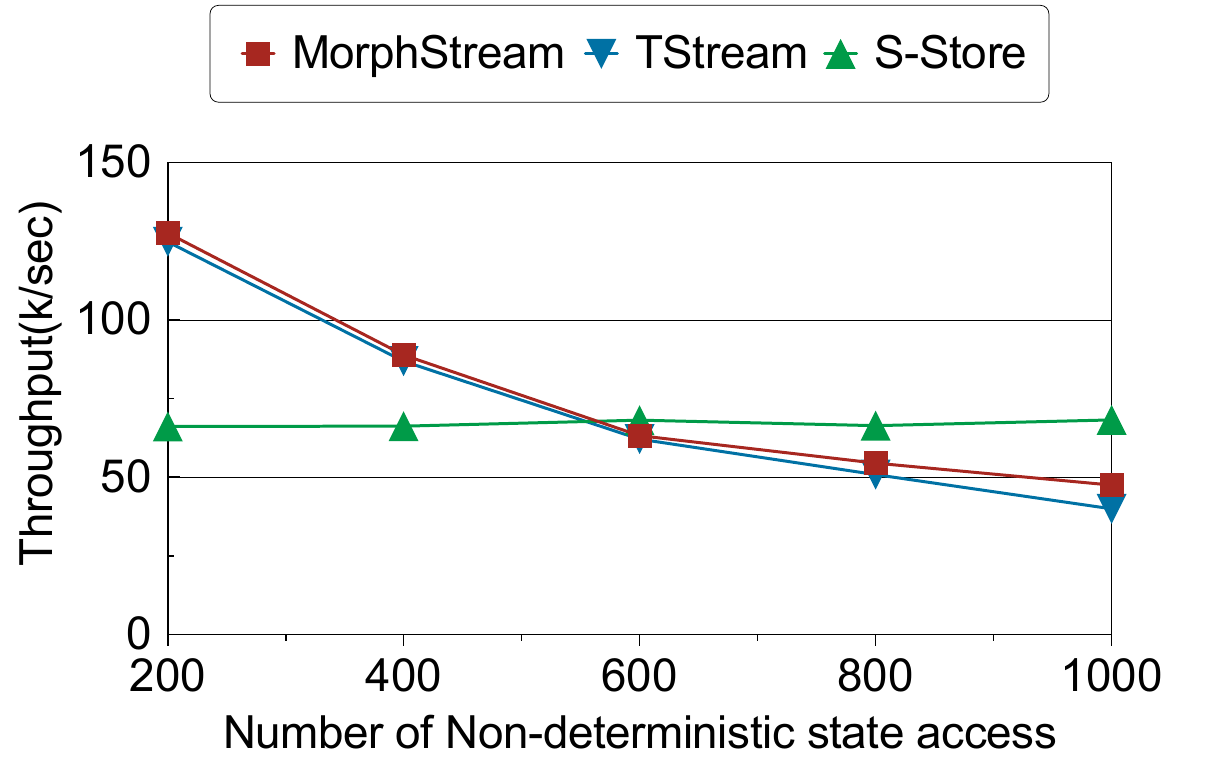}
	\caption{Evaluation of Non-deterministic Queries.}
        \label{fig:non_exp}
	\end{minipage}
\end{figure}
\myc{
We implemented the \textit{GrepSum with non-deterministic queries} to demonstrate \system's support for non-deterministic state access. It involves processing state transactions that read specific states and compute summation results, which are then written back to a designated state. However, the state to be accessed can be deterministic or non-deterministic. Deterministic state access is based solely on the input event, while non-deterministic state access depends on additional factors such as user-defined functions, timers, or random values~\cite{Clonos}. 
}

\myc{
In this experiment, we focused on tuning the number of state transactions involving non-deterministic state access to investigate its impact on the performance of \system. The results, shown in Figure~\ref{fig:non_exp}, led to two key observations. First, the number of non-deterministic state accesses had no significant effect on the performance of \sstore. This is because \sstore executes dependent operations sequentially, resulting in minimal overhead for handling non-deterministic state access. Second, both \system and \tstream experienced notable performance degradation as the number of non-deterministic state accesses increased. This can be attributed to the higher \tpg planning overhead associated with a large number of non-deterministic state accesses, requiring the addition of virtual operations and tracking dependencies across all operation chains. The results of this experiment indicate a significant optimization space within \system to enable efficient tracking of dependencies for non-deterministic state access operations. For instance, one potential approach could involve predicting the accessed state and conducting pilot runs of non-deterministic state access operations. However, developing an efficient and low-overhead training model for this purpose poses challenges and remains a topic for future work.
}

\subcompact
\begin{figure}[t]
	\begin{minipage}{0.5\textwidth}
	\centering
	\subfloat[Runtime Breakdown]{
        \includegraphics[width=0.79\textwidth]{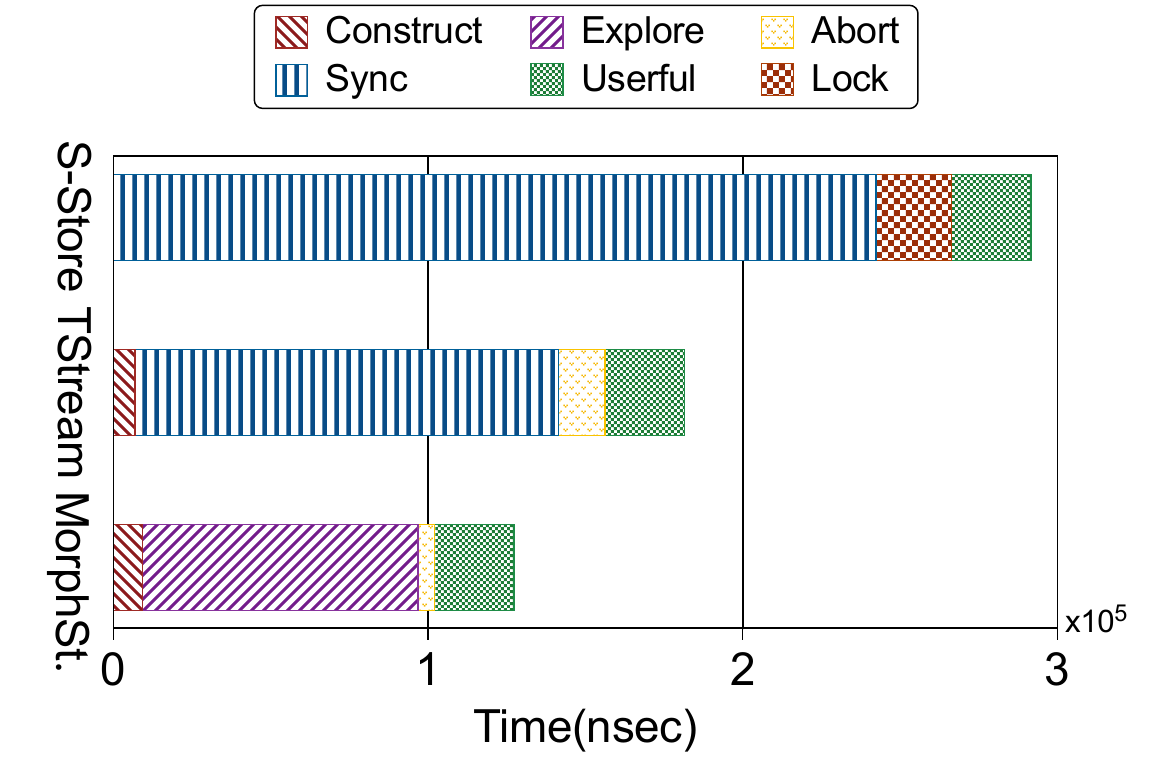}
			\label{fig:Breakdown}
		}	
  
        \subfloat[Memory Footprint]{
		\includegraphics[width=0.79\textwidth]{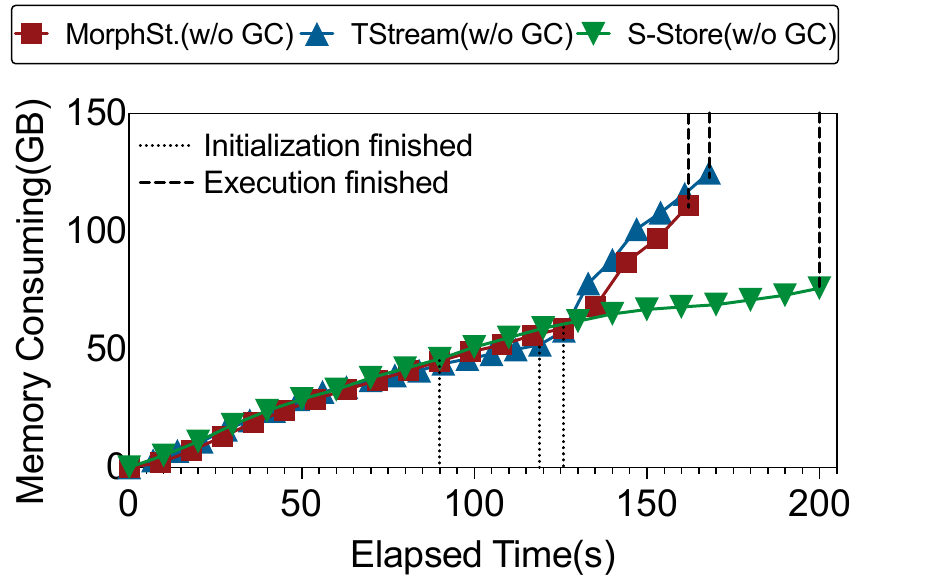}   
		\label{fig:memory_footprint}
	}
	\end{minipage}
	\caption{System overhead.}
    \label{fig:system_overhead}
\end{figure}

\begin{figure}[t]
\begin{center}
	{\setlength{\fboxsep}{1pt}
			     \hspace{0.99em}			     
				\begin{minipage}{0.5\textwidth}
					\includegraphics[width=\columnwidth]{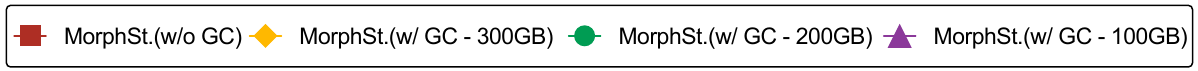}
				\end{minipage}
		}
\end{center}  
	\begin{minipage}{0.5\textwidth}
	\centering 
	\subfloat[Throughput]{
        \includegraphics[width=0.79\textwidth]{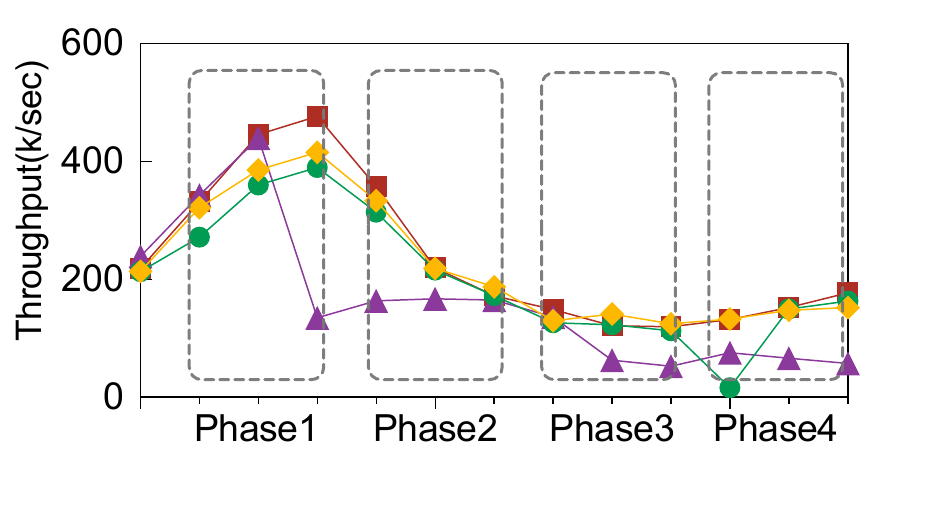}
			\label{fig:jvm_throught}
		}	
  
        \subfloat[Memory Footprint]{
		\includegraphics[width=0.79\textwidth]{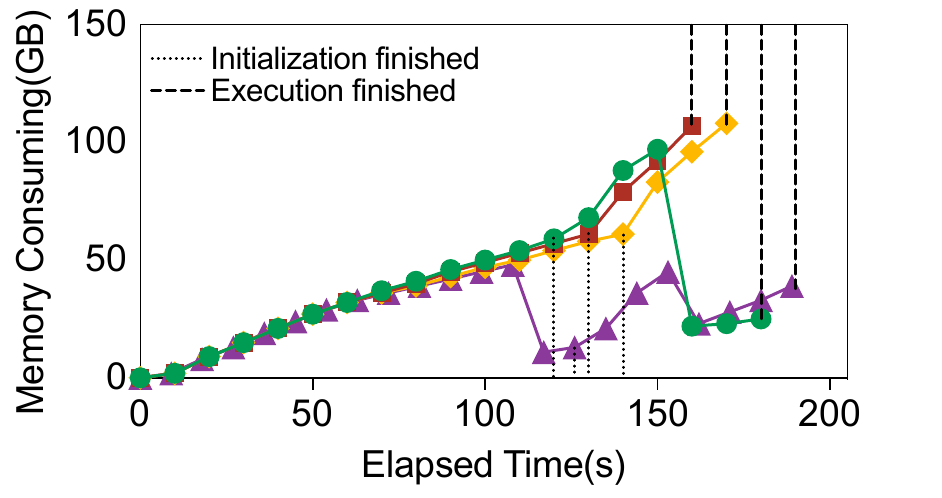}   
		\label{fig:jvm_memory_footprint}
	}
	\end{minipage}
	\caption{Impact of clean-up under varying JVM size.}
    \label{fig:varying_jvm}
\end{figure}

\begin{figure}[t]
\begin{center}
		{\setlength{\fboxsep}{1pt}
			\hspace{0.99em}			 
				\begin{minipage}{0.3\textwidth}
					\includegraphics[width=\columnwidth]{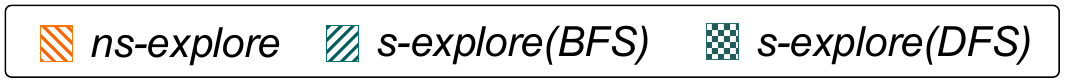}
				\end{minipage}
		}
\end{center}	
	\begin{minipage}{0.5\textwidth}
 \centering
	\subfloat[Varying Punc. Interval]{
			\includegraphics[width=0.79\textwidth]{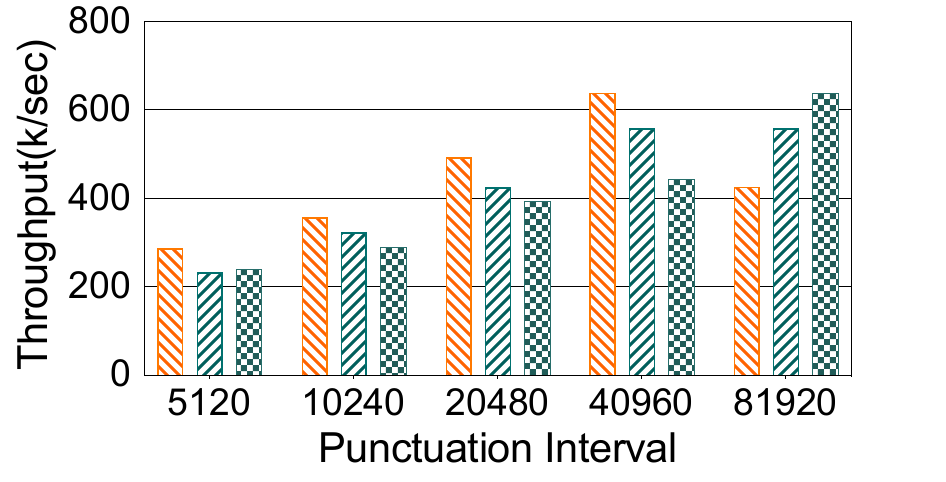}  
			\label{fig:low_skewness}
		}
  
        \subfloat[Varying Workload Skewness]{
		\includegraphics[width=0.79\textwidth]{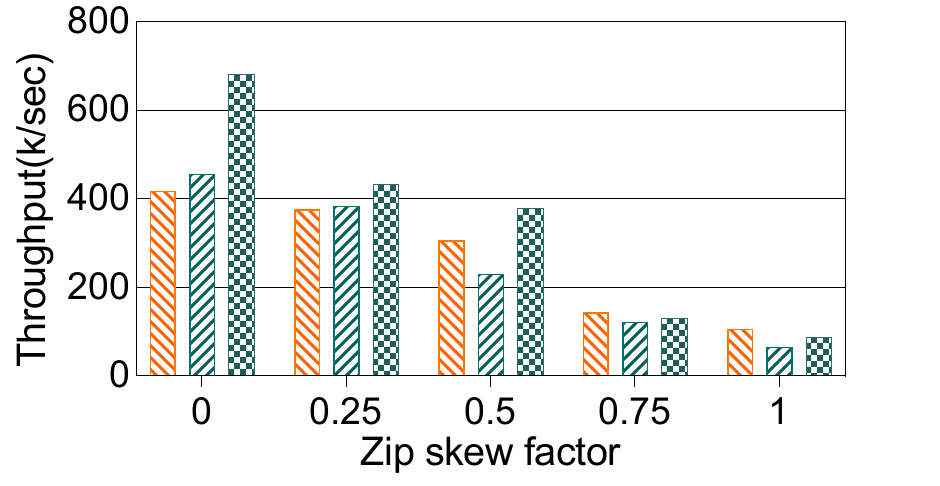}   
		\label{fig:state_access_skewness}
	}
        \end{minipage}
	\caption{Exploration strategy decision.}
    \label{figures:exploration}
\end{figure}

\subcompact
\subsection{Overhead}
\label{subsec:overhead}
\system achieves adaptive scheduling at the cost of more complex runtime operations such as data structures constructing and exploring available state access operations. 
These extra operations can negatively impact the system in the following two ways: 1) the complex construction and exploration process may increase the latency of transaction processing, and 2) the auxiliary data structure will increase the memory consumption of the application. 

\subcompact
\subsubsection{Latency overhead} 
Following a prior work~\cite{tstream}, we show the time breakdown in the following aspects. 
1) \emph{Useful Time} refers to the time spent on doing actual work including accessing shared mutable states and performing user-defined functions.
2) \emph{Sync Time} refers to the time spent on synchronization, including blocking time before lock insertion is permitted or blocking time due to synchronization barriers during mode switching.
3) \emph{Lock Time} refers to the time spent on inserting locks after it is permitted.
4) \emph{Construct Time} refers to the time spent on constructing the auxiliary data structures, e.g., \tpg in \system and operation chains in \tstream.
5) \emph{Explore Time} refers to the time spent on exploring available operations to process.
6) \emph{Abort Time} refers to the wasted computation time due to abort and redos.

Figure~\ref{fig:Breakdown} shows the time breakdown when the system runs the dynamic workload in Section~\ref{subsec:evaluation}. There are three key takeaways.
First, although \tstream and \system spend a significant portion of time during construction (\emph{Construct Time}), they successfully reduce synchronization (\emph{Sync Time}) and lock (\emph{Lock Time}) overhead compared to \sstore. This explains their better performance on multicore processors.
Second, \tstream has the highest abort time (\emph{Abort Time}) because \tstream has to redo the entire batch of transactions when a transaction abort happens. 
In contrast, \sstore spends little time in abort as it involves little redo of state transactions because dependent transactions are executed serially.
Third, we can see that \system still spends a significant fraction of time performing exploration (\emph{Explore Time}). This is mainly caused by excessive message-passing among threads. In the future, we plan to investigate more efficient exploration strategies such as prioritizing mechanisms~\cite{kwok1999static} in \system.

\subsubsection{Memory footprint} 
\label{subsec:memory-footprint}
We use the dynamic workload in Section~\ref{subsec:evaluation} to evaluate memory footprint. Figure~\ref{fig:memory_footprint} demonstrates the memory consumption of three TSPEs without GC involved. 
In the initialization stage, the memory consumption of all systems is almost the same. 
\system spends more time during initialization compared to \tstream as it needs to initialize more data structures to support adaptive scheduling. 
During runtime, \system and \tstream consume a similar amount of memory per batch of state transactions, and both consume much more than \sstore. This is because they construct auxiliary data structures for scheduling, and especially they may maintain multiple physical copies of each state at different timestamps during execution (Section~\ref{subsec:abort_handling}).
Note that, as we have configured \system to not clear temporal objects and the JVM size to be large enough (300GB), the total memory usage keeps increasing until execution is finished, during which no GC is triggered. 
We plan to incorporate stream compression~\cite{Cstream,CompressStreamDB} in \system to reduce such high memory footprints in future.

\subsubsection{Clean-up and GC overhead}
\label{subsec:GC}
Figure~\ref{fig:varying_jvm} shows the impact of clean-up under varying JVM sizes from 100GB to 300GB. We can see that enabling clear temporal objects brings down the performance of \system up to 12.8\%, and still outperforms TStream and S-Store. In Figure~\ref{fig:varying_jvm}(b), the memory usage fluctuates up and down when the JVM size is set to 100GB or 200 GB because the JVM periodically reclaims (GC) the temporary objects in the continued processing of data streams.
\subcompact
\begin{figure*}
\begin{center}
		{\setlength{\fboxsep}{1pt}
			\hspace{0.99em}
			 \vspace{-0.3cm}
				\begin{minipage}{0.3\textwidth}
					\includegraphics[width=\columnwidth]{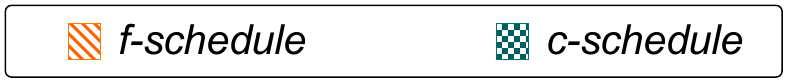}
				\end{minipage}
		}
\end{center}
	\centering
	\begin{minipage}{\textwidth}
	\subfloat[Cyclic/Acyclic]{
	       	\includegraphics[width=0.33\textwidth]{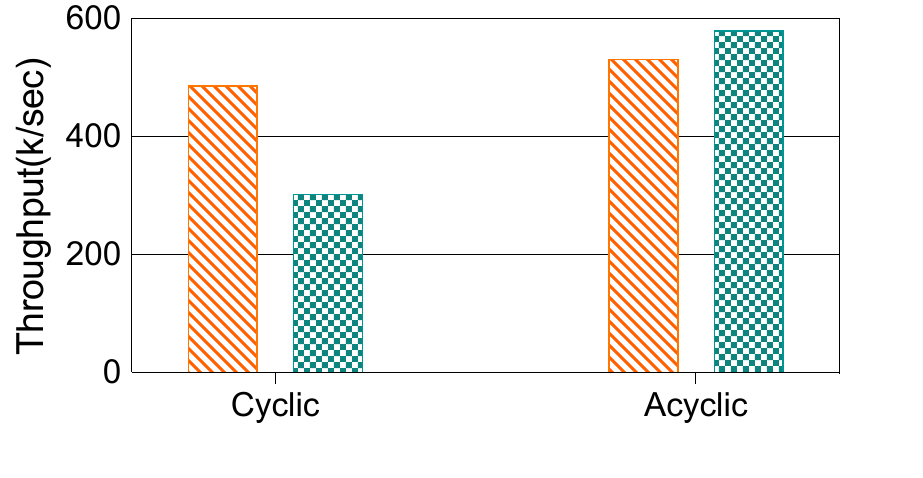}
            \label{fig:low_num_access}
        }	
	\subfloat[Varying Punctuation Interval]{
		\includegraphics[width=0.33\textwidth]{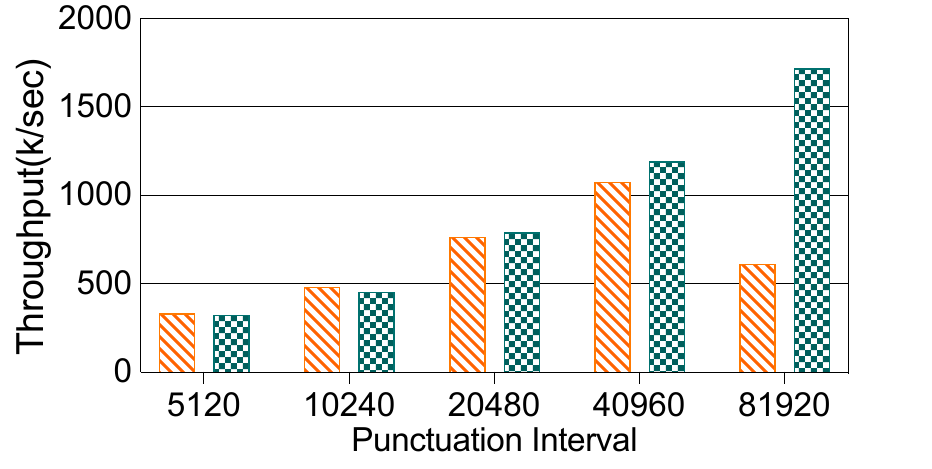}
        \label{fig:punctuation_interval}
		}	
    \subfloat[Varying Ratio of Multi-Accesses]{   
		\includegraphics[width=0.33\textwidth]{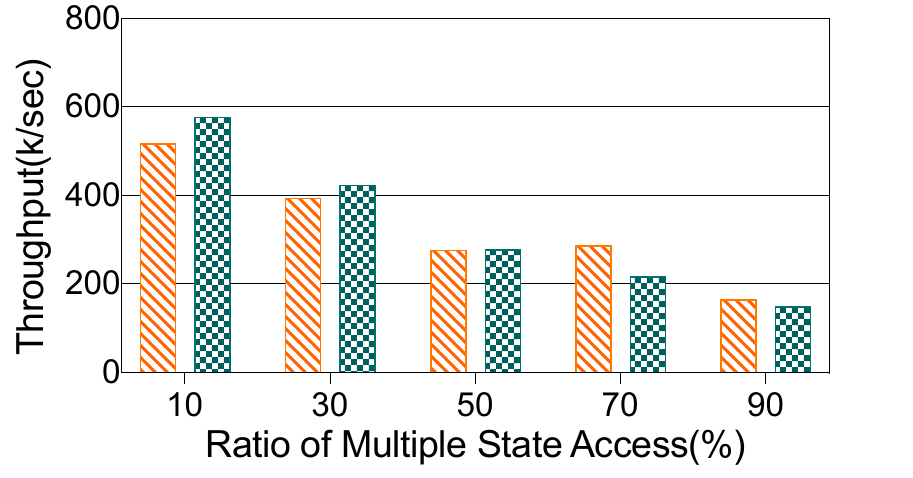}   
		\label{fig:state_access_number}
	}
	\end{minipage}
	\caption{Granularity decision.}
    \label{figures:Granularity decision}
\end{figure*}
\begin{figure}
\begin{center}
		{\setlength{\fboxsep}{1pt}
			\hspace{0.99em}
				\begin{minipage}{0.2\textwidth}
					\includegraphics[width=\columnwidth]{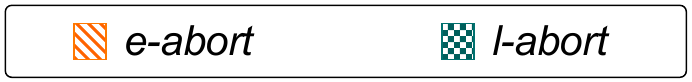}
				\end{minipage}
		}
    \end{center}
	\begin{minipage}{0.5\textwidth}
 	\centering
        \subfloat[Varying Compute Complex.]{   
		\includegraphics[width=0.79\textwidth]{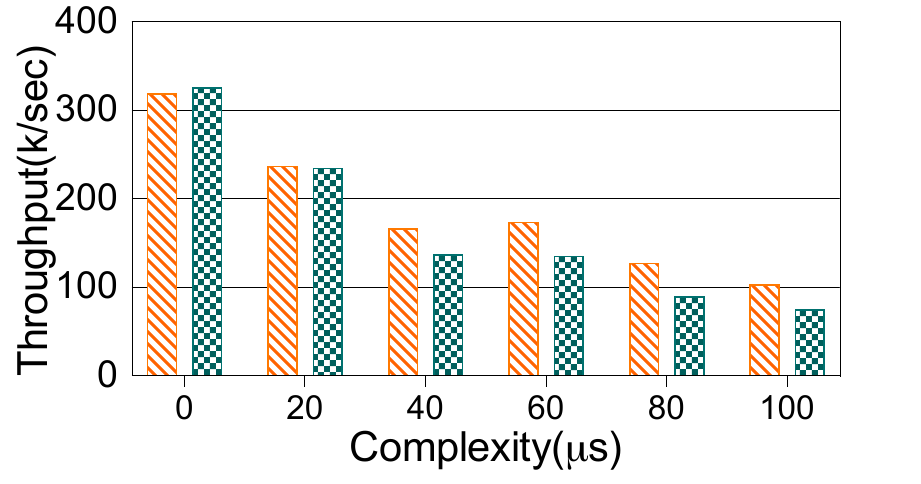}   
		\label{fig:high_abort_ratio}
	}
 
	\subfloat[Varying Abort Ratio]{
		\includegraphics[width=0.79\textwidth]{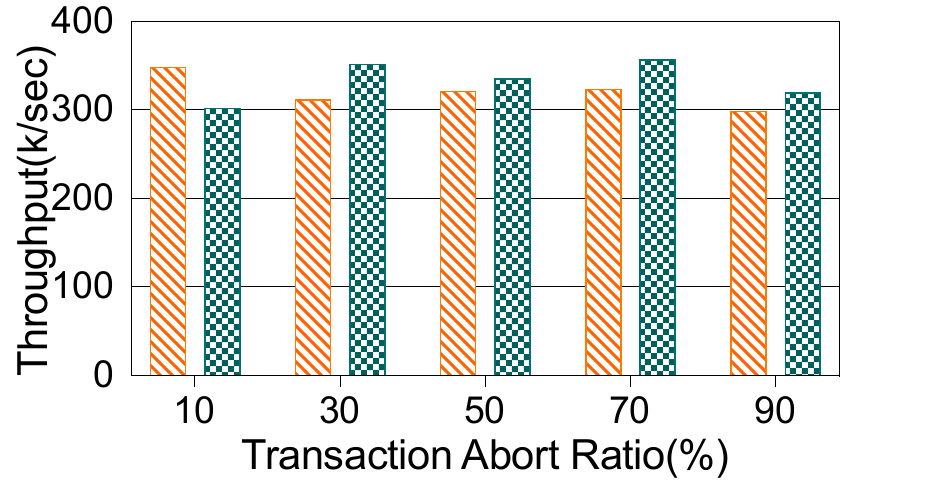}
        \label{fig:abort_ratio}
		}	
	\end{minipage}
	\caption{Abort handling decision.}
 \end{figure}
 
\subcompact
\subsection{Impact of Scheduling Decisions}
\label{subsec:scheduling_decision}
In this section, we evaluate the impact of varying scheduling decisions under different workload characteristics using GS due to its flexibility.

\subsubsection{Impact of Exploration Strategies}
We first study the effectiveness of different exploration strategies (i.e., \nse vs. \se) mainly affected by the \emph{punctuation interval} and the \emph{workload skewness}, as discussed in Section~\ref{subsec:model}.
Figure~\ref{fig:low_skewness} shows the effects of selecting different exploration strategies under varying \emph{punctuation interval} and low \emph{workload skewness}.
\nse works better when \emph{punctuation interval} is low, while \se works better when \emph{punctuation interval} is high. 
This is due to the linear proportionality between the \emph{punctuation interval} and the number of dependencies (\tds/\pds) of the constructed \tpg.
When the \emph{punctuation interval} is low, \nse resolves the rare dependencies as soon as an operation has been successfully processed, leading to higher system concurrency. \se works better otherwise as the notification overhead of the \nse approach keeps increasing with more dependencies in the workloads. However, \se has a constant construction and synchronization overhead for dependencies resolution.
Figure~\ref{fig:state_access_skewness} shows the effects of selecting different exploration strategies under varying \emph{workload skewness} and high \emph{punctuation interval}. We can see that \se works better when the state accesses are uniformly distributed, i.e., the Zipf skew factor is 0. \nse works better when the state accesses are skewed. This is because a skewed workload leads to load unbalance among threads and intensifies the synchronization overhead when \se is applied as summarized in Table~\ref{tab:decisions}.
\subsubsection{Impact of Scheduling Granularities}
In this section, we study the effectiveness of different scheduling granularities (i.e., \fsu v.s. \csu), which are affected by the following key workload characteristics, namely \emph{cyclic/acyclic}, \emph{number of state access}, \emph{punctuation interval}, and the \emph{ratio of multi-accesses}.
First, 
Figure~\ref{fig:low_num_access} shows the results of different scheduling granularities under the workload with or without cyclic dependencies. \csu performs better when there is no cyclic dependency among the batched scheduling units since each thread can schedule a group of operations together as a scheduling unit to amortize the context switching overheads. 
However, our further experiments reveal that when there is a large number of state access, \fsu is always better than \csu, regardless of whether there are circular dependencies. This is mainly due to the fact that even without circular dependencies, a large number of state accesses will increase the number of \pd, causing a significant overhead on resolving the dependencies among operations of the same group.
Second,
Figure~\ref{fig:punctuation_interval} shows how varying \emph{punctuation interval} affects the selection of scheduling unit granularities when there are no cyclic dependencies. 
We set the number of state accesses to one to avoid the effect of \pd, so the punctuation interval only controls the number of \tds in the \tpg.
We can see that \csu achieves higher throughput at higher punctuation intervals. 
When the punctuation interval is high, the large number of \tds increases the context-switching overhead in \fsu, which is why the performance of \fsu decreases when the punctuation interval is large, such as 81920.
In contrast, \csu schedules the operations in a group resulting in lower context-switching overhead on resolving \td compared to the \fsu.
Third,
Figure~\ref{fig:state_access_number} shows that \fsu works better when the ratio of multiple state access is high, while \csu works better when the ratio is low. The ratio of multiple state access controls the number of \pds among operations, as we can see that the \pd affects the performance of \csu significantly. This is mainly because the execution concurrency drops when the number of \pds is high. 
 
  \begin{figure}
    \label{figures:abort}
	\centering
	\begin{minipage}{0.5\textwidth}
 \centering
	\subfloat[Micro-architectural Analysis]{
		\includegraphics[width=0.79\textwidth]{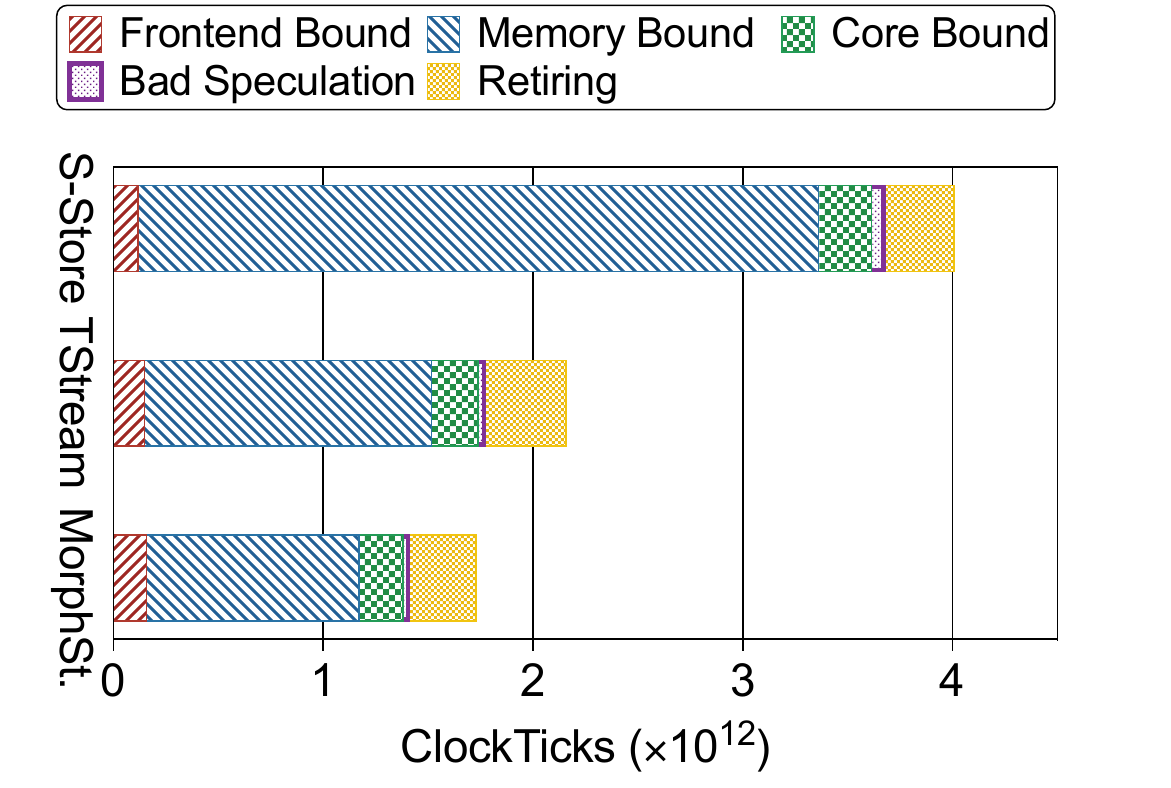}
        \label{fig:uarch_topdown}
	}
 
        \subfloat[Scalability Comparison]{
		\includegraphics[width=0.79\textwidth]
		{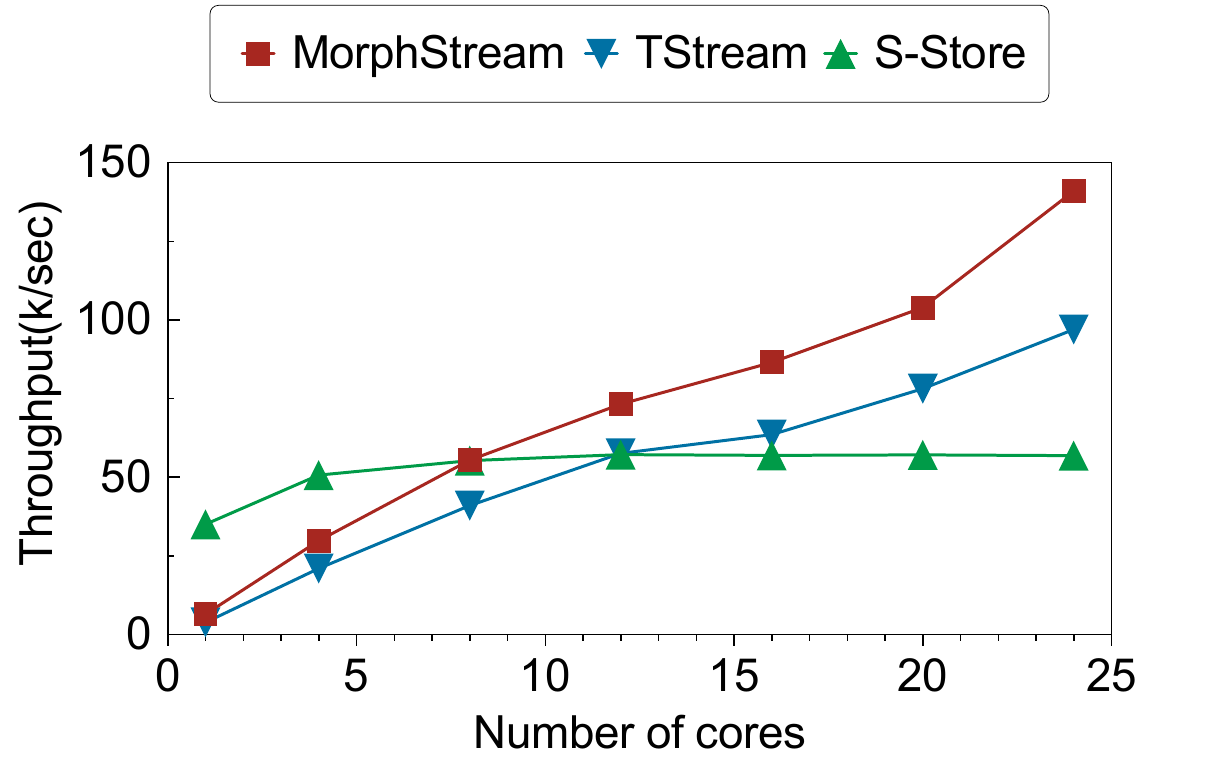}   
		\label{fig:multi_core}
	}
 	\end{minipage}
	\caption{Impact of Modern Hardware.}
    \label{figures:hardware}
\end{figure}

\subsubsection{Impact of Abort Handling Mechanisms}
Finally, we study the impact of two abort handling mechanisms (i.e., \ea v.s. \la) mainly affected by the \emph{abort ratio} and the \emph{computation complexity} workload characteristics.
Figure~\ref{fig:high_abort_ratio} shows the comparison results of varying \emph{computation complexity} when the \emph{abort ratio} is high. 
A lower computation complexity leads to a low redo overhead, and \la handles frequent aborts together to reduce the context switching overheads. 
\ea is better otherwise, as it makes a minimum impact on the ongoing execution of other operations.
Figure~\ref{fig:abort_ratio} shows the results of different abort handling mechanisms under different abort ratios when the computation complexity is low. The results indicate that as the ratio of aborting transactions increases, \la works better. The key reason is that when the computation complexity is low, the context-switching overheads and the synchronization overhead among threads to achieve fine-grained state rollback and redo become the major bottlenecks.
\subcompact

\subcompact
\subsection{Impact of Modern Hardware}
\label{subsec:modern_hardware}
In this section, we compare \system with existing TSPEs on how they interact with modern multicore processors from the modern hardware architecture perspective. 

\textbf{Micro-architectural Analysis.}
We take SL as an example to show the breakdown of the execution time according to the Intel Manual. Figure~\ref{fig:uarch_topdown} compares the time breakdown of different TSPEs. We measure the hardware performance counters through Intel Vtune Profiler during the algorithm execution and compute the top-down metrics. 
We have three major observations.\margi{R5D8}First, the breakdown results reaffirm our previous analysis that \system spends up to 2.3x fewer clock ticks for transaction processing compared to \tstream and \sstore, because of its more efficient adaptive scheduling strategies. Second, all three TSPEs \change{are Memory Bounded}, i.e., a large proportion of CPU cycles are spent due to memory access instructions: \system (58.5\%), \tstream (63.3\%), and \sstore (80.9\%).
The detailed profiling with Intel Vtune Profiler reveals that it is commonly due to the heavy usage of latches to resolve dependencies among transactions while accessing the shared-mutable state. Both \tstream and \sstore have a higher Memory Bound than \system due to the higher synchronization cost. Nevertheless, \change{Figure~\ref{fig:uarch_topdown} and\margii{-15pt}{R5D8}Figure~\ref{fig:Breakdown} jointly indicate that \system can adopt more efficient exploration strategies to further improve its performance.}

\textbf{Multicore Scalability.}
Figure~\ref{fig:multi_core} shows the scalability comparison among TSPEs, with two major observations. 
First, \system outperforms prior schemes with an increasing number of cores confirming the good scalability of \system. However, there is still a large room for further improving \system towards linearly scale-up, the reason being that it becomes memory bounded as Figure~\ref{fig:uarch_topdown} previously shown.
Second, when the number of cores is low, \system performs even worse than \sstore due to the large constant overhead of \tpg construction process. In a resource constraint setting, existing non-adaptive solutions may be more favoured.

\subcompact
\subcompact
\subsection{Case Study}
\label{subsec:use_cases}
\myc{
We further demonstrate the practical usefulness of \system with two case studies that need to maintain shared states with the requirements of high performance and correctness.
}

\myc{
\subsubsection{Online Social Event Detection}
}
\myc{
Detecting unexpected data patterns on social media platforms is a pressing need known as online social event detection (OSED)~\cite{fedoryszak2019real,hasan2018survey}. This process entails continually processing streaming data while reading and updating three shared states: \textit{Word}, which represents meaningful concepts in alphabetic form; \textit{Tweet}, which denotes a standardized string of words; and \textit{Cluster}, which is a collection of tweets encoding similar event information.
}

\begin{figure}[t]
    \includegraphics[width=0.95\linewidth]{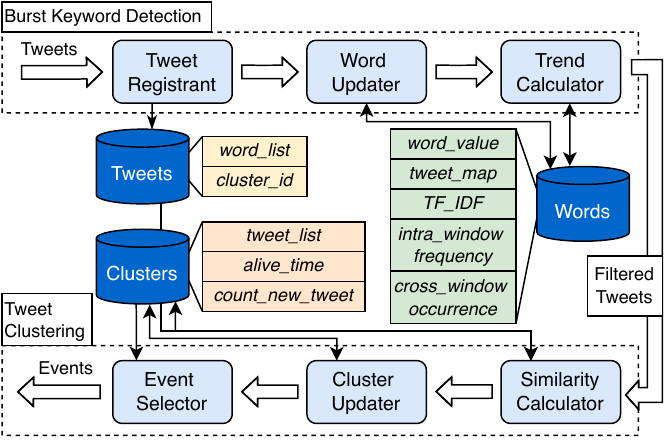}
    \caption{Workflow of Online Social Event Detection.}
    \label{fig:Online_Event_Detection_Workflow}
\end{figure}

\myc{
Numerous ways have been proposed for OSED~\cite{hasan2018survey}. In this demonstration, we use a novel method called \textit{Hybrid Event Detection}~\cite{sahin2019streaming}, which consists of two stages:
\emph{(1) Burst Keyword Detection:}
This stage finds and labels words with a high frequency growth across windows as \textit{burst keywords}. 
\emph{(2) Tweet Clustering:}
Tweets containing \textit{burst keywords} are grouped together into tweet clusters based on cosine similarity. At the end of each window, clusters with significant growth rates are identified as output events.
}

\myc{
OSED exhibits several distinguishing features:
\emph{Firstly,} high data processing capacity is required to successfully capture trending topics from enormous streams of social media posts, along with low latency to respond to real-time event detection requests.
\emph{Secondly,} it requires highly concurrent accesses to shared states (i.e., \textit{Word}, \textit{Tweet}, and \textit{Cluster}) to track event evolution, making it difficult to assure state consistency in complex transactional dependencies.
\emph{Lastly,} the workload characteristics of the input social media post stream are highly dynamic, posing a challenge in allocating tasks effectively among executors.
}

\myc{
To address these challenges, \system structures the three types of shared state (\textit{Word}, \textit{Tweet}, and \textit{Cluster}) as shared mutable key-value pairs, as depicted in Figure~\ref{fig:Online_Event_Detection_Workflow}. Each pair contains multiple fields facilitating the necessary computations for event detection. To ensure state consistency, \system maps simultaneous state access operations as a single transaction. It effectively resolves transaction dependencies and adapts to the most suitable scheduling strategy under dynamic workload characteristics.
}

\myc{
The workflow of OSED on \system is illustrated in Figure~\ref{fig:Online_Event_Detection_Workflow}.
The \textit{Tweet Registrant} initiates the process by registering pre-processed tweets into the state, decomposing them into word tokens, and distributing them downstream. The \textit{Word Updater} then updates the words' frequencies to the state. The \textit{Trend Calculator} then identifies burst keywords with significant increases in TF-IDF values across windows and emits tweets with these burst keywords for clustering. These actions are carried out after the status of all words in the current window has been updated, ensuring that no burst keywords are overlooked, as controlled by the \textit{punctuation} mechanism~\cite{mao2023morphstream}. After receiving filtered tweets, the \textit{Similarity Calculator} compares them to existing clusters and determines the most suitable clusters. If no matching clusters are found, new clusters are created. Once all tweets in the window are merged into clusters by the \textit{Cluster Updater}, the \textit{Event Selector} identifies clusters with high growth rates as events and propagates them as output. Meanwhile, clusters with no updates over an extended period of time are eliminated.
}

\begin{figure}[t]
    \centering
    \includegraphics[width=0.95\linewidth]{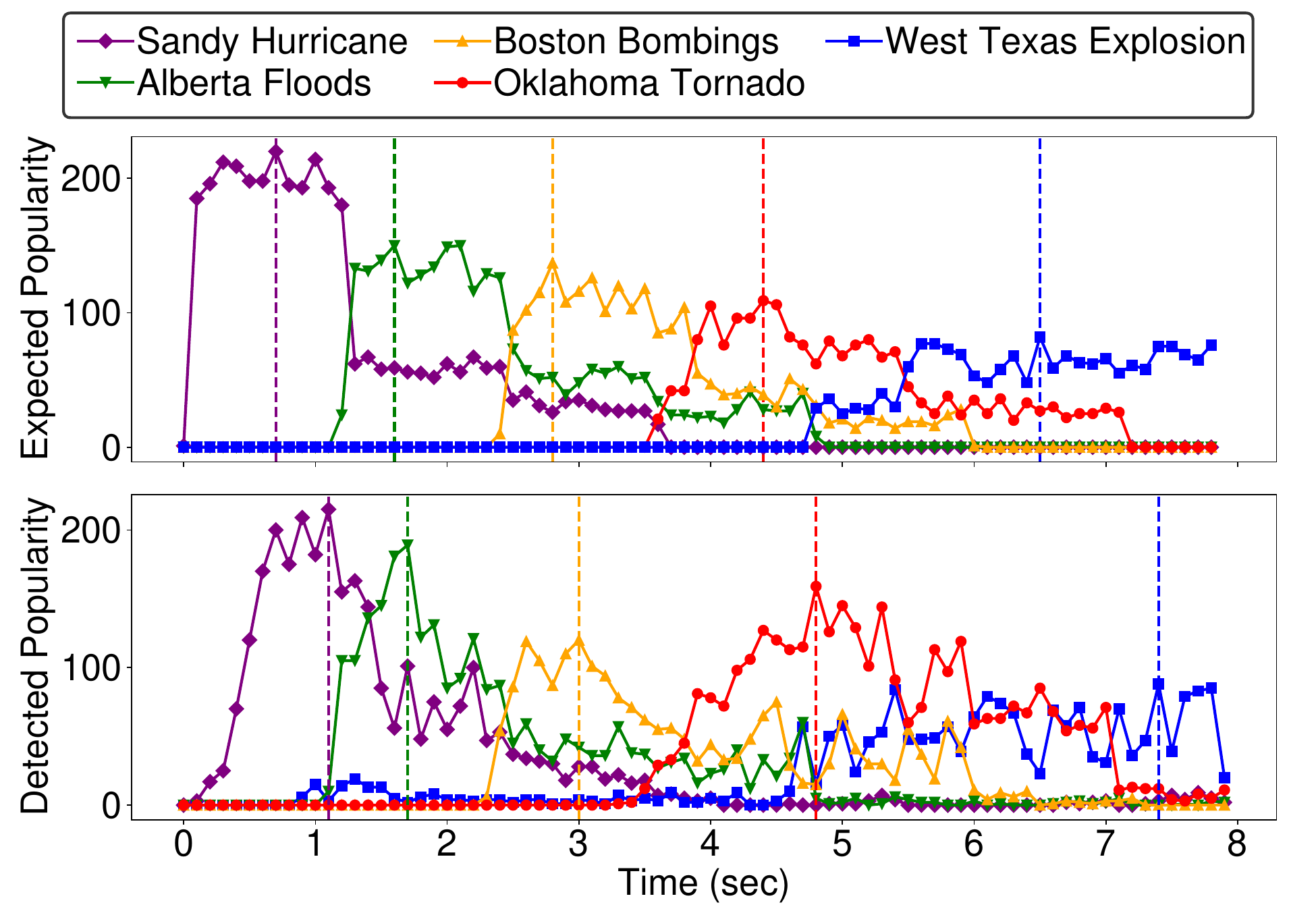}
    \caption{Event popularity over time - expected and detected event evolution.}
    \label{fig:ed_performance}
\end{figure}

\myc{
We conducted our analysis using a dataset of real-world tweets~\cite{olteanu2014crisislex}, comprising English tweets from five crisis events that occurred in the United States between 2012 and 2013. This dataset consists of approximately 30,000 tweets, both event-related and non-event-related. For our application, we deployed four executors for each operator, with a total punctuation interval of 400 tweets. Each thread was allocated 100 tweets per batch to execute.
In Figure~\ref{fig:ed_performance}, we present the performance results of our online event detection implemented using \system, compared to the actual evolution of events over time. The popularity of events is measured by the number of new tweets merged into a specific event cluster within each time window. 
The results demonstrate that our online event detection, supported by \system, accurately detects the emergence of events and effectively captures changes in event popularity trends within seconds, as indicated by the time difference in event popularity summits. We also observed that \system achieved an overall throughput of up to 1.3k tweets per second for processing and detecting events. These findings provide compelling evidence of \system's ability to efficiently support complex real-time applications.
}

\subsubsection{Real-time Stock Exchange Analysis}
One common stock exchange analysis (SEA)~\cite{sse} task is to get the turnover rates of stocks by calculating the trade ratio of each stock for every period of time. 
The input data is a stream of quotes and a stream of trades containing the trade results of matching quotes.
The query joins the traded stream and the quotes stream ($S$) over the same stock id within the same period of time (i.e., window). 
Specifically, given the unbounded nature of streaming data, windows are commonly employed to restrict the number of tuples involved in the computation. 
Subsequently, window join matches the traded and quote records for each stock to calculate its associated turnover rates.
The SEA can be implemented using the hash-based window join algorithm.
The algorithm maintains two hash tables, one for each input stream.
When it receives a tuple from Traded (or Quote) stream, it inserts the tuple into the hash table of Traded (or Quote) and immediately probes the hash table of the opposite stream Quote (or Traded). 

\begin{figure}[t]
    \includegraphics[width=0.95\linewidth]{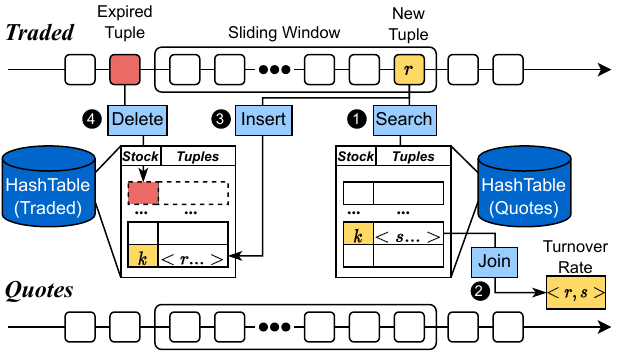}
    \caption{Workflow of Stock Exchange Turnover Rates Analysis using Hash-based  Window Join.}
    \label{fig:Stock_Exchange_Workflow}
\end{figure}


\myc{
With \system, we can intuitively map the hash table structure to the shared state and model insertion and probe requests to the hash table as state transactions. 
An overview of the stock analysis workflow implemented by hash-based window join is shown in Figure~\ref{fig:Stock_Exchange_Workflow}. 
\system maintains two hash tables \(Index(Traded)\) and \(Index(Quotes)\) for streams \(Traded\) and \(Quotes\), respectively as shared state. 
The key $k$ of the state is the stock id and the value $<r, ...>$ contains all arrived tuples in the current window slide.
}

\myc{
The join operation involves the following steps. 
When a new tuple \(r\) arrives from stream \(Traded\), \system searches for matching tuples in \(HashTable(Quotes)\) by efficient index lookup. Once matched tuples \(<s, ...>\) are identified, \system calculates turnover rates accordingly during a window slide and propagates the result $\texttt{<}r,s\texttt{>}$ as the join output. 
Subsequently, it deletes the expired tuple and inserts the new tuple \(r\) into \(HashTable(Traded)\), updating the multi-version state storage accordingly. 
All accesses to hash tables, such as insert and delete, are mapped to state access operations and are subsequently modelled as state transactions in \system. Only when all operations of a state transaction are performed successfully, the transaction is committed. Otherwise, the state is restored to the latest version before abort, and the transaction will be re-executed.
}


\begin{figure}[t]
    \centering
    \includegraphics[width=0.8\linewidth]{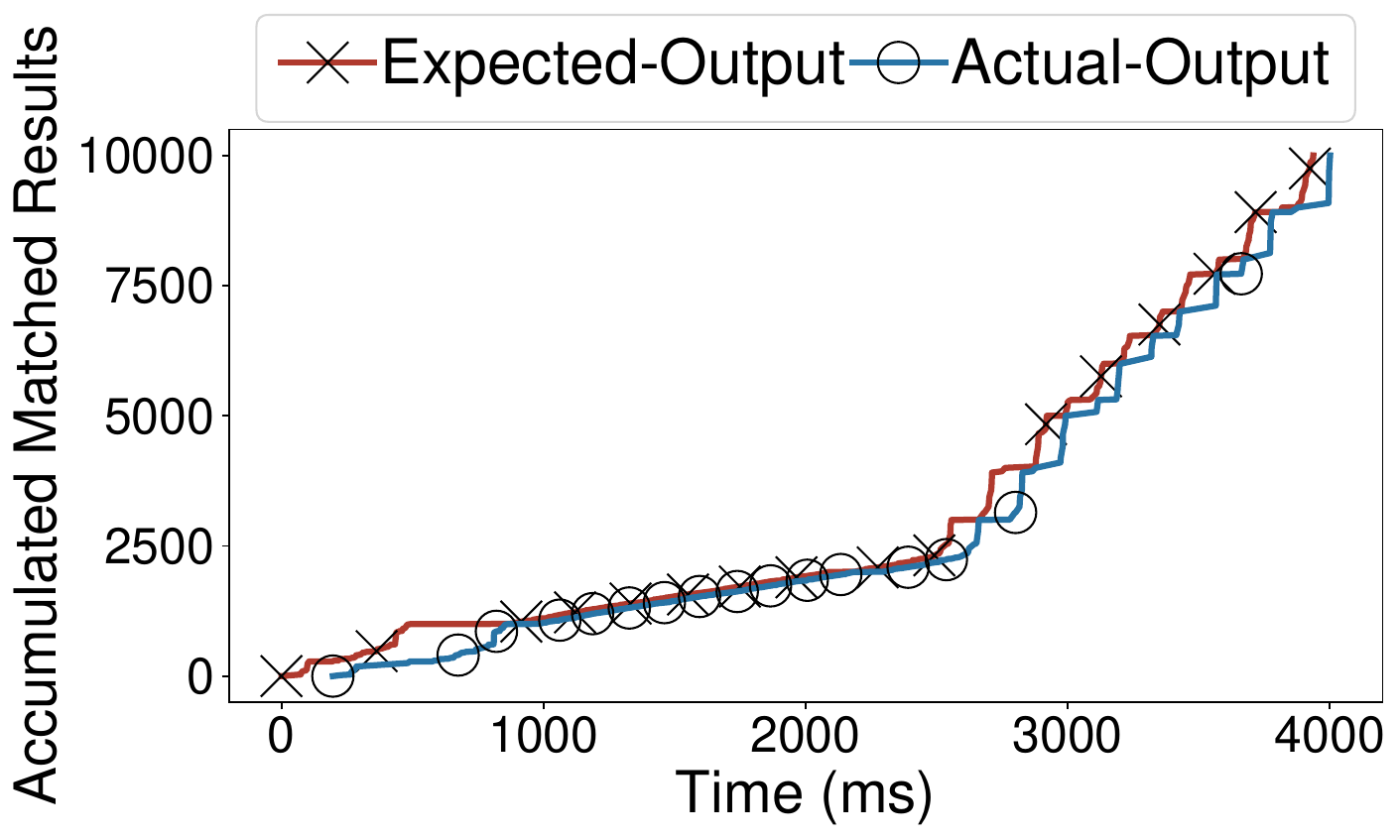}
    \caption{Stock exchange accumulated matched results - expected and actual output.}
    \label{fig:stock_performance}
\end{figure}


\myc{
Figure~\ref{fig:stock_performance} shows the performance results of the stock exchange analysis implemented based on \system.
For evaluation, we utilized a real-world stock exchange dataset~\cite{sse} containing tens of millions of quote and traded records. 
The application was deployed with 10 executors, and the batch interval was configured as 1k, where each executor will be evenly allocated with 1k records per batch for execution. 
The primary focus of the performance evaluation is on the expected accumulated matched result generated by traded/quote events and the actual result output by \system.
The performance results demonstrate that \system consistently outputs the actual results within the millisecond level.
We also measured that the throughput of the \system can process up to 70k events per second.
These findings confirm the efficiency of implementing real-time financial applications in \system, as it can achieve the ACID guarantees for transactions while maintaining high throughput and low latency.
}
 
\compact
\section{Related Work}
\label{sec:related}
Unlike key-value stores, database systems, or conventional SPEs, TSPEs such as Streaming-Ledger~\cite{Transactions2018}, \sstore~\cite{S-Store}, FlowDB~\cite{Affetti:2017:FIS:3093742.3093929}, \tstream~\cite{tstream}, and \system are based on a unique computational model, where each input tuple from data streams may involve multiple keys. Thus the processing of tuples can lead to potentially conflicting shared mutable state accesses. 
Such a unique system feature has been originally motivated by a list of stream applications~\cite{ACEP,Botan09} and is applied or encouraged to be applied in emerging use case scenarios~\cite{meehan2017data,Transactions2018,10.1007/978-3-030-19274-7_10}. 
Some of the novel design challenges and optimization opportunities of TSPEs have been discussed in previous works~\cite{S-Store,tstream}. 
The experimental results showed previously in Figure~\ref{fig:overview_comparison} also confirm that conventional SPEs can not efficiently handle the targeted applications of TSPEs.

Executing each state transaction one by one following the event sequence naturally leads to the correct schedule but seriously limits system concurrency~\cite{ACEP}. 
Recent works have proposed adopting partitioning and decomposition to optimize the performance of transaction processing, such as~\cite{Bernstein:1999:CCS:337919.337922,Recovery,Shasha:1995:TCA:211414.211427,BengChinTKDE16}. Similar ideas have also been adopted in TSPEs. 
For example, S-Store~\cite{S-Store} adopts state partitioning with extensions of guaranteeing state access ordering~\cite{S-Store-demo}, while TStream~\cite{tstream} adopts transaction decomposition to improve multicore scalability further. 
However, each existing system is designed with a non-adaptive scheduling strategy and favours a subset of workload characteristics.
\system deviates from existing solutions. 
It explores fine-grained workload characteristics of every batch of state transactions. It then makes the correct scheduling based on a decision model to morph the current scheduling strategy into a better-performing strategy.

Despite the large body of research on the scheduling problem in a general context~\cite{hall1997scheduling,T-storm,briskstream}, task scheduling for TSPEs presents subtle but unique requirements~\cite{10.1145/167088.167254,Robert2011,hou1994genetic,669967}, largely due to the integrated stream processing and transactional contexts~\cite{tstream}. For instance, 
the scheduling unit can be reconfigured by the system, e.g., by splitting state transactions into operations and regrouping by keys. It is thus difficult (if not impossible) to quantitatively model the objective function of scheduling plans in TSPEs. We hence propose to model the scheduling of TSPEs into a three-dimensional scheduling decision problem and guide it with a heuristic-based decision model. Furthermore, the scheduling overhead is now on the critical path, prohibiting any sophisticated optimization algorithms. 
\compact
\section{Conclusion}
\label{sec:conclusion}
In this work, we introduced \system, a TSPE designed to optimize parallelism and performance for stream applications managing shared mutable states. 
Through a unique three-stage execution paradigm, \system enables dynamic scheduling and parallel processing in TSPEs. 
The \emph{Planning Stage} effectively tracks dependencies using a two-phase \tpg construction process and virtual operations. 
The \emph{Scheduling Stage} dynamically adjusts exploration strategy, granularity, and abort handling, while \emph{Execution Stage} ensures correct transaction handling using a Stateful \tpg. 
Our experiment showcased the remarkable capabilities of \system, significantly outperforming existing TSPEs in various scenarios.
We plan to continue improving \system, with a focus on integrating fault tolerance mechanisms and addressing other challenges presented by evolving real-time stream applications.

\compact
\begin{acknowledgements}
We would like to express our gratitude to Mr.Siqi Xiang for his invaluable contributions to this work, including data collection, preprocessing, and assistance with some experiments.
This work is supported by the National Research Foundation, Singapore and Infocomm Media Development Authority under its Future Communications Research \& Development Programme FCP-SUTD-RG-2022-005, and a SUTD Start-up Research Grant (SRT3IS21164). 
Haikun Liu's work is supported by the National Natural Science Foundation of China under grant No.62072198. 
Volker Markl's work is supported by the DFG Priority Program (MA4662-5), the German Federal Ministry of Education and Research (BMBF) under grants 01IS18025A (BBDC - Berlin Big Data Center) and 01IS18037A (BIFOLD - Berlin Institute for the Foundations of Learning and Data). This work was also supported in part by AMD under the Heterogeneous Accelerated Compute Clusters (HACC) program (formerly known as the XACC program - Xilinx Adaptive Compute Cluster program).
\end{acknowledgements}

\bibliographystyle{spmpsci}
\bibliography{mylib}

\begin{thebibliography}{10}
\providecommand{\url}[1]{{#1}}
\providecommand{\urlprefix}{URL }
\expandafter\ifx\csname urlstyle\endcsname\relax
  \providecommand{\doi}[1]{DOI~\discretionary{}{}{}#1}\else
  \providecommand{\doi}{DOI~\discretionary{}{}{}\begingroup
  \urlstyle{rm}\Url}\fi

\bibitem{sse}
Shanghai stock exchange, \url{http://english.sse.com.cn/} (2018).
\newblock Last Accessed: 2020-06-29

\bibitem{Affetti:2017:FIS:3093742.3093929}
Affetti, L., Margara, A., Cugola, G.: Flowdb: Integrating stream processing and
  consistent state management.
\newblock In: Proceedings of the 11th ACM International Conference on
  Distributed and Event-based Systems, Debs '17, pp. 134--145. Acm, New York,
  NY, USA (2017).
\newblock \doi{10.1145/3093742.3093929}.
\newblock \urlprefix\url{http://doi.acm.org/10.1145/3093742.3093929}

\bibitem{Arasu:2004:LRS:1316689.1316732}
Arasu, A., Cherniack, M., Galvez, E., Maier, D., Maskey, A.S., Ryvkina, E.,
  Stonebraker, M., Tibbetts, R.: Linear road: A stream data management
  benchmark.
\newblock In: Proceedings of the Thirtieth International Conference on Very
  Large Data Bases - Volume 30, Vldb '04, pp. 480--491. VLDB Endowment (2004).
\newblock \urlprefix\url{http://dl.acm.org/citation.cfm?id=1316689.1316732}

\bibitem{Bernstein:1999:CCS:337919.337922}
Bernstein, A.J., et~al.: Concurrency control for step-decomposed transactions.
\newblock Inf. Syst. 1999 \textbf{24}(9), 673--698 (1999).
\newblock \urlprefix\url{http://dl.acm.org/citation.cfm?id=337919.337922}

\bibitem{Botan09}
Botan, I., Cho, Y., Derakhshan, R., Dindar, N., Haas, L.M., Kim, K., Lee, C.,
  Mundada, G., Shan, M.C., Tatbul, N., et~al.: Design and implementation of the
  maxstream federated stream processing architecture.
\newblock Technical Report/ETH Zurich, Department of Computer Science
  \textbf{632} (2009)

\bibitem{Botan12}
Botan, I., Fischer, P.M., Kossmann, D., Tatbul, N.: Transactional stream
  processing.
\newblock In: Proceedings of the 15th International Conference on Extending
  Database Technology, EDBT '12, p. 204–215. Association for Computing
  Machinery, New York, NY, USA (2012).
\newblock \doi{10.1145/2247596.2247622}.
\newblock \urlprefix\url{https://doi.org/10.1145/2247596.2247622}

\bibitem{carbone2015apache}
Carbone, P., Katsifodimos, A., Ewen, S., Markl, V., Haridi, S., Tzoumas, K.:
  Apache flink: Stream and batch processing in a single engine.
\newblock The Bulletin of the Technical Committee on Data Engineering
  \textbf{38}(4) (2015)

\bibitem{flink}
Carbone, P., Katsifodimos, A., Ewen, S., Markl, V., Haridi, S., Tzoumas, K.:
  Apache flink (2018).
\newblock \urlprefix\url{http://flink.apache.org/}

\bibitem{S-Store-demo}
Cetintemel, U., et~al.: S-store: A streaming newsql system for big velocity
  applications.
\newblock Proc. VLDB Endow. 2014 \textbf{7}(13), 1633--1636 (2014).
\newblock \doi{10.14778/2733004.2733048}.
\newblock \urlprefix\url{http://dx.doi.org/10.14778/2733004.2733048}

\bibitem{dynamicWorkload}
Ding, B., Chaudhuri, S., Gehrke, J., Narasayya, V.: Dsb: a decision support
  benchmark for workload-driven and traditional database systems.
\newblock Proceedings of the VLDB Endowment \textbf{14}(13), 3376--3388 (2021)

\bibitem{fedoryszak2019real}
Fedoryszak, M., Frederick, B., Rajaram, V., Zhong, C.: Real-time event
  detection on social data streams.
\newblock In: Proceedings of the 25th ACM SIGKDD international conference on
  knowledge discovery \& data mining, pp. 2774--2782 (2019)

\bibitem{10.1145/167088.167254}
Feldmann, A., Kao, M.Y., Sgall, J., Teng, S.H.: Optimal online scheduling of
  parallel jobs with dependencies.
\newblock In: Proceedings of the Twenty-Fifth Annual ACM Symposium on Theory of
  Computing, STOC '93, p. 642–651. Association for Computing Machinery, New
  York, NY, USA (1993).
\newblock \doi{10.1145/167088.167254}.
\newblock \urlprefix\url{https://doi.org/10.1145/167088.167254}

\bibitem{golab2}
Golab, L., Bijay, K.G., {\"O}zsu, M.T.: On concurrency control in sliding
  window queries over data streams.
\newblock In: Y.~Ioannidis, M.H. Scholl, J.W. Schmidt, F.~Matthes,
  M.~Hatzopoulos, K.~Boehm, A.~Kemper, T.~Grust, C.~Boehm (eds.) Advances in
  Database Technology - EDBT 2006, pp. 608--626. Springer Berlin Heidelberg,
  Berlin, Heidelberg (2006)

\bibitem{hall1997scheduling}
Hall, L.A., Schulz, A.S., Shmoys, D.B., Wein, J.: Scheduling to minimize
  average completion time: Off-line and on-line approximation algorithms.
\newblock Mathematics of operations research \textbf{22}(3), 513--544 (1997)

\bibitem{hasan2018survey}
Hasan, M., Orgun, M.A., Schwitter, R.: A survey on real-time event detection
  from the twitter data stream.
\newblock Journal of Information Science \textbf{44}(4), 443--463 (2018)

\bibitem{he2008mars}
He, B., et~al.: Mars: a mapreduce framework on graphics processors.
\newblock In: PACT'08, pp. 260--269. IEEE (2008)

\bibitem{hou1994genetic}
Hou, E.S., Ansari, N., Ren, H.: A genetic algorithm for multiprocessor
  scheduling.
\newblock IEEE Transactions on Parallel and Distributed systems \textbf{5}(2),
  113--120 (1994)

\bibitem{669967}
Kwok, Y.K., Ahmad, I.: Benchmarking the task graph scheduling algorithms.
\newblock In: Proceedings of the First Merged International Parallel Processing
  Symposium and Symposium on Parallel and Distributed Processing, pp. 531--537
  (1998).
\newblock \doi{10.1109/IPPS.1998.669967}

\bibitem{kwok1999static}
Kwok, Y.K., Ahmad, I.: Static scheduling algorithms for allocating directed
  task graphs to multiprocessors.
\newblock ACM Computing Surveys (CSUR) \textbf{31}(4), 406--471 (1999)

\bibitem{mao2023morphstream}
Mao, Y., Zhao, J., Zhang, S., Liu, H., Markl, V.: Morphstream: Adaptive
  scheduling for scalable transactional stream processing on multicores.
\newblock In: Proceedings of the 2023 International Conference on Management of
  Data (SIGMOD), SIGMOD '23. Association for Computing Machinery, New York, NY,
  USA (2023)

\bibitem{meehan2017data}
Meehan, J., Aslantas, C., Zdonik, S., Tatbul, N., Du, J.: Data ingestion for
  the connected world.
\newblock In: CIDR (2017)

\bibitem{S-Store}
Meehan, J., Tatbul, N., Zdonik, S., Aslantas, C., Cetintemel, U., Du, J.,
  Kraska, T., Madden, S., Maier, D., Pavlo, A., Stonebraker, M., Tufte, K.,
  Wang, H.: S-store: Streaming meets transaction processing.
\newblock Proc. VLDB Endow. \textbf{8}(13), 2134–2145 (2015).
\newblock \doi{10.14778/2831360.2831367}.
\newblock \urlprefix\url{https://doi.org/10.14778/2831360.2831367}

\bibitem{nikolov2006graph}
Nikolov, N.S., Tarassov, A.: Graph layering by promotion of nodes.
\newblock Discrete Applied Mathematics \textbf{154}(5), 848--860 (2006)

\bibitem{olteanu2014crisislex}
Olteanu, A., Castillo, C., Diaz, F., Vieweg, S.: Crisislex: A lexicon for
  collecting and filtering microblogged communications in crises.
\newblock In: Proceedings of the AAAI Conference on Weblogs and Social Media
  (ICWSM '14). AAAI Press, Ann Arbor, MI, USA (2014)

\bibitem{Robert2011}
Robert, Y.: Task Graph Scheduling, pp. 2013--2025.
\newblock Springer US, Boston, MA (2011).
\newblock \doi{10.1007/978-0-387-09766-4_42}.
\newblock \urlprefix\url{https://doi.org/10.1007/978-0-387-09766-4_42}

\bibitem{sahin2019streaming}
Sahin, O., KARAGOZ, P., Tatbul, N.: Streaming event detection in microblogs:
  Balancing accuracy and performance.
\newblock In: International Conference on Web Engineering, pp. 123--138.
  Springer (2019).
\newblock \doi{10.1007/978-3-030-19274-7_10}

\bibitem{10.1007/978-3-030-19274-7_10}
Sahin, O.C., Karagoz, P., Tatbul, N.: Streaming event detection in microblogs:
  Balancing accuracy and performance.
\newblock In: M.~Bakaev, F.~Frasincar, I.Y. Ko (eds.) Web Engineering, pp.
  123--138. Springer International Publishing, Cham (2019)

\bibitem{Shasha:1995:TCA:211414.211427}
Shasha, D., et~al.: Transaction chopping: Algorithms and performance studies.
\newblock ACM Trans. Database Syst. 1995 \textbf{20}(3), 325--363 (1995).
\newblock \doi{10.1145/211414.211427}.
\newblock \urlprefix\url{http://doi.acm.org/10.1145/211414.211427}

\bibitem{Clonos}
Silvestre, P.F., Fragkoulis, M., Spinellis, D., Katsifodimos, A.: Clonos:
  Consistent causal recovery for highly-available streaming dataflows.
\newblock In: Proceedings of the 2021 International Conference on Management of
  Data, pp. 1637--1650 (2021)

\bibitem{bidding}
Tan, J., Zhong, M.: An online bidding system (obs) under price match mechanism
  for commercial procurement.
\newblock Applied Mechanics and Materials, 2014 \textbf{556}, 6540--6543
  (2014).
\newblock \doi{10.4028/www.scientific.net/AMM.556-562.6540}

\bibitem{storm}
Toshniwal, A., Taneja, S., Shukla, A., Ramasamy, K., Patel, J.M., Kulkarni, S.,
  Jackson, J., Gade, K., Fu, M., Donham, J., et~al.: Apache storm (2018).
\newblock \urlprefix\url{http://storm.apache.org/}

\bibitem{Transactions2018}
Transactions, S.A., Data, S.: {Data Artisans Streaming Ledger Serializable ACID
  Transactions on Streaming Data,
  \url{https://www.da-platform.com/streaming-ledger}}  (2018)

\bibitem{Tucker:2003:EPS:776752.776780}
Tucker, P.A., et~al.: Exploiting punctuation semantics in continuous data
  streams.
\newblock TKDE'03 \textbf{15}(3), 555--568 (2003).
\newblock \doi{10.1109/TKDE.2003.1198390}.
\newblock \urlprefix\url{http://dx.doi.org/10.1109/TKDE.2003.1198390}

\bibitem{ACEP}
Wang, D., Rundensteiner, E.A., Ellison III, R.T.: Active complex event
  processing over event streams.
\newblock Proc. VLDB Endow. \textbf{4}(10), 634--645 (2011).
\newblock \doi{10.14778/2021017.2021021}.
\newblock \urlprefix\url{http://dx.doi.org/10.14778/2021017.2021021}

\bibitem{Recovery}
Wu, Y., Guo, W., Chan, C.Y., Tan, K.L.: Fast failure recovery for main-memory
  dbmss on multicores.
\newblock In: Proceedings of the 2017 ACM International Conference on
  Management of Data, SIGMOD ’17, p. 267–281. Association for Computing
  Machinery, New York, NY, USA (2017).
\newblock \doi{10.1145/3035918.3064011}.
\newblock \urlprefix\url{https://doi.org/10.1145/3035918.3064011}

\bibitem{T-storm}
Xu, J., Chen, Z., Tang, J., Su, S.: T-storm: Traffic-aware online scheduling in
  storm.
\newblock In: Proceedings of the 2014 IEEE 34th International Conference on
  Distributed Computing Systems, Icdcs '14, pp. 535--544. IEEE Computer
  Society, Washington, DC, USA (2014).
\newblock \doi{10.1109/icdcs.2014.61}.
\newblock \urlprefix\url{http://dx.doi.org/10.1109/ICDCS.2014.61}

\bibitem{BengChinTKDE16}
Yao, C., Agrawal, D., Chen, G., Lin, Q., Ooi, B.C., Wong, W.F., Zhang, M.:
  Exploiting single-threaded model in multi-core in-memory systems.
\newblock IEEE Transactions on Knowledge and Data Engineering \textbf{28}(10),
  2635--2650 (2016).
\newblock \doi{10.1109/TKDE.2016.2578319}

\bibitem{spark}
Zaharia, M., Das, T., Li, H., Hunter, T., Shenker, S., Stoica, I.: Discretized
  streams: Fault-tolerant streaming computation at scale.
\newblock In: Proceedings of the twenty-fourth ACM symposium on operating
  systems principles, pp. 423--438 (2013)

\bibitem{Cstream}
Zeng, X., Zhang, S.: Parallelizing stream compression for iot applications on
  asymmetric multicores.
\newblock In: 2023 IEEE 39th International Conference on Data Engineering
  (ICDE). IEEE (2023)

\bibitem{briskstream}
Zhang, S., et~al.: Briskstream: Scaling data stream processing on shared-memory
  multicore architectures.
\newblock In: SIGMOD '19, SIGMOD '19, pp. 705--722. ACM, New York, NY, USA
  (2019).
\newblock \doi{10.1145/3299869.3300067}.
\newblock \urlprefix\url{http://doi.acm.org/10.1145/3299869.3300067}

\bibitem{tstream}
Zhang, S., Wu, Y., Zhang, F., He, B.: Towards concurrent stateful stream
  processing on multicore processors.
\newblock In: 2020 IEEE 36th International Conference on Data Engineering
  (ICDE), pp. 1537--1548. IEEE (2020)

\bibitem{CompressStreamDB}
Zhang, Y., Zhang, F., Li, H., Zhang, S., Du, X.: Compressstreamdb: Fine-grained
  adaptive stream processing without decompression.
\newblock In: 2023 IEEE 39th International Conference on Data Engineering
  (ICDE). IEEE (2023)

\end{thebibliography}
\end{document}